\documentclass[aip,jcp,amsmath,amssymb,preprint,graphicx]{revtex4-2}

\usepackage{dcolumn}
\usepackage{bm}
\usepackage[usenames,dvipsnames]{color}

\usepackage{graphicx}

\usepackage{hyperref}
\hypersetup{
colorlinks=true,
linkcolor=blue,
citecolor=blue,
filecolor=green,
urlcolor=blue,
}

\newcommand{\threejm}[6]{ \left(\begin{array}{ccc} #1 & #3 & #5\\
                                              #2 & #4 & #6
                                \end{array}
                          \right)}

\usepackage{color}
\definecolor{darkblue}{rgb}{0.0,0.0,0.5}
\definecolor{Blue2}{rgb}{0.,0.25,0.7}

\definecolor{crimson}{rgb}{0.80,0.05,0.15}

\begin{document}


\title{Rigorous quantum calculations for atom-molecule chemical reactions in electric fields: from single to multiple partial wave regimes}




\author{Timur V. Tscherbul}
\affiliation{Department of Physics, University of Nevada, Reno, NV, 89557, USA}\email[]{ttscherbul@unr.edu}
\author{Roman V. Krems}
\affiliation{Department of Chemistry, University of British Columbia, Vancouver, BC, V6T 1Z1, Canada}\email[]{rkrems@ubc.ca}


\date{\today}

\begin{abstract}\small
We present an efficient method for rigorous quantum calculations of cross sections for atom-molecule reactive scattering in the presence of a dc electric field.  
The wavefunction of the  reaction complex is expanded in an overcomplete set of arrangement-dependent Fock-Delves hyperspherical basis functions and the interactions of the reactants and products  with electric fields are  accounted for in the total angular momentum representation.
 A significant computational challenge affecting our previously developed approach [Phys. Rev. Lett. {\bf 115}, 023201 (2015)]  is addressed by an efficient asymptotic frame transformation between the hyperspherical and Jacobi coordinates in the presence of an external field.
 Using accurate {\it ab initio} potential energy surfaces,   we calculate total and state-resolved cross sections for the chemical reactions LiF$(v=1,j=0)$ + H $\to$ Li + HF($v'=0,j'$) and F + HD$(v=0,j=0)$ $\to$ HF + D, DF~+~H as functions of collision energy and electric field strength. The field dependence of the cross sections for the LiF~+~H chemical reaction exhibits resonance structure mediated by tunneling-driven interactions between reactants and products.  No significant field effects are found for the F + HD $\to$ HF~+~D, DF~+~H chemical reaction at 1 Kelvin, even for state-resolved transitions and with  field magnitudes reaching 200 kV/cm.
 Our calculations illustrate the essential role of basis set convergence for the proper interpretation of external field effects on chemical reaction dynamics.  
 While reduced-basis calculations for the F~+~HD reaction  indicate significant effects of electric fields on product state distributions, these effects vanish when the number of total angular momentum basis states is increased.  
\end{abstract}\normalsize

\maketitle 


\section{Introduction}
\label{sec:intro}

Controlling chemical reactions by external fields has long been a major goal in several research fields, driving advances in coherent control of molecular dynamics \cite{Shapiro:12}, experiments with decelerated \cite{Meerakker:12,Vogels:15,Jongh:20,Tang:23} and merged   \cite{Lavert-Ofir:14,Klein:16,Margulis:22,Margulis:23} molecular beams, ultracold chemistry \cite{Krems:08,Balakrishnan:16,Bohn:17,Liu:22,Karman:24}, and spin echo studies of molecule - surface interactions \cite{Godsi:17}.  It has been argued  \cite{Krems:08} that molecular encounters can be most easily modified by external fields at low temperatures, where perturbations induced by external fields exceed the kinetic energy of molecular collisions.
In the last decade, major advances in molecular cooling experiments have resulted in diverse arrays of trapped polar molecules, ranging from ultracold alkali metal dimers \cite{Bohn:17} and laser-coolable molecules CaH \cite{VazquezCarson:22}, CaF \cite{Truppe:17,Anderegg:18}, SrF \cite{Barry:14},  YO \cite{Burau:23,Langen:24} to polyatomic molecules such as  CH$_3$ \cite{Yang:17}, CH$_3$F \cite{Zeppenfeld:12,Prehn:16}, SrOH \cite{Kozyryev:17}, CaOH \cite{Anderegg:23}, and CaOCH$_3$ \cite{Mitra:20}.
These experiments provide a new paradigm for exploring  control of chemical reactivity in the predominantly quantum regime.

Experimental studies of atom-molecule and molecule-molecule collisions at
 ultracold ($T\lesssim 1$~mK)  temperatures \cite{Krems:08,Balakrishnan:16,Bohn:17,Karman:24} have demonstrated extensive control over scattering observables (such as collision cross sections and reaction rates)  near magnetic Feshbach  resonances  
\cite{Yang:19,Wang:21,Park:23,Park:23b,Karman:23,Morita:24}. Inelastic losses in ultracold molecular gases can now be effectively suppressed by engineering long-range repulsive barriers via electric and/or microwave shielding \cite{Gorshkov:08,Karman:18,Lassabliere:18,Anderegg:21}, a strategy that led to the observation of electric-field-induced resonant shielding in KRb~+~KRb collisions \cite{Matsuda:20},
 the creation of long-range field-linked  (NaK)$_2$ bound states \cite{Chen:23} and the realization of a Bose-Einstein condensate of polar NaCs molecules \cite{Bigagli:24}.
 It has also been demonstrated that  quantum coherence \cite{Devolder:20,Devolder:21,Devolder:23,Luke:24} and entanglement of reactants \cite{Devolder:25} can be exploited for controlling chemical reactivity at ultralow temperatures. Remarkably, quantum coherence in the nuclear spin degrees of freedom has been shown to persist throughout the ultracold KRb~+~KRb chemical reaction \cite{Liu:24}.

While these quantum effects can be leveraged to achieve robust control over bimolecular chemical reactions \cite{Krems:08,Balakrishnan:16,Liu:22},  
they can only be harnessed at low and ultralow temperatures, where quantum scattering is dominated by a single partial wave or a small number of partial waves \cite{Krems:05,Krems:08,Carr:09}. At higher temperatures, these phenomena are expected to be gradually washed out by thermal randomization of molecular collisions, leading to partial wave scrambling due to incoherent addition of multiple partial wave contributions \cite{Devolder:23}. 
However, several observations suggest that some mechanisms or regimes of quantum reactive scattering may remain sensitive to external fields at elevated temperatures. 

First, the differences between the triplet and singlet phase shifts can become  synchronized over a wide range of partial waves,  a phenomenon known as partial wave phase locking \cite{Sikorsky:18,Cote:18,Tomza:19,Katz:25,Deiss:24}. This can lead to pronounced quantum interference effects  \cite{Sikorsky:18,Devolder:23} and magnetic Feshbach resonances  \cite{Walewski:25} that  persist far into the multiple partial wave regime.
Second, in the studies of collision stereodynamics,  significant changes in scattering cross sections  (up to $40-50$\%) have been observed upon switching the orientation of the diatomic molecular axis with respect to the incident velocity vector at collision energies as high as  500 cm$^{-1}$ for   NO~+~Ar collisions\cite{Nichols:15} and $\simeq$4,000~cm$^{-1}$  for the chemical reaction H~+~HD $\to$ H$_2$~+~D     \cite{Wang:23}.
At  these elevated collision energies, the steric effects can be largely understood in terms of  the initial molecular bond axis orientation  in the laboratory frame \cite{Nichols:15,McCrea:25,Wang:23}, i.e., without explicitly including external fields in quantum dynamical calculations.  
 Finally, specific mechanisms of chemical reactions can potentially remain sensitive to external fields even at ambient temperatures. For example, the tunneling-driven reaction of F$(^2P_{3/2})$ with H$_2$ may be affected by small admixtures of spin-orbit excited states of F introduced by couplings induced by an external magnetic field \cite{Krems:08}.  Similarly, an interplay of external field-dependent couplings and fine intra-molecular interactions can modify the relative spin orientation of reacting open-shell molecules, thereby affecting the non-adiabatic dynamics of a chemical process \cite{Tscherbul:06,Abrahamsson:07}. It is not yet known if the response of such reaction dynamics to external fields survives thermal averaging.

The remarkable success of experiments with ultracold molecules along with the open challenge for controlling chemical reactivity at higher temperatures require the development of rigorous quantum theory for efficient calculations of state-resolved probabilities of reactive scattering in external fields. For ultracold scattering, the current key goal is to extend rigorous calculations to heavy diatomic molecules and to polyatomic molecules with a large density of quantum states participating in reaction dynamics. 
Recent progress of theoretical work is marked by  converged coupled-channel  calculations of hyperfine-to-rotational energy transfer in ultracold Rb~+~KRb collisions including the spin degrees of freedom of KRb and Rb, along with both the intramolecular and intermolecular hyperfine interactions and external magnetic fields \cite{Liu:25}. These calculations were performed using the rigid-rotor approximation and hence neglected the vibrational degrees of freedom of the Rb-KRb complex and the conical intersection (CI) between relevant potential energy surfaces (PESs). Non-adiabatic effects have been considered in quantum reactive scattering calculations of the K~+~KRb $\to$ K$_2$~+~Rb chemical reaction \cite{Silva:25}. These calculations treated the CI and rovibrational degrees of freedom of the K-KRb reaction complex exactly but did not account for  the spin degrees of freedom, hyperfine interactions and external fields.

At present, chemically reactive collisions in the presence of external fields remain out of reach of rigorous quantum scattering theory for most of the molecular species cooled and trapped in recent experiments \cite{Langen:24}, even in the $s$-wave limit of ultracold chemistry.  
As noted above, the current key goal for ultracold chemistry  is to extend reactive quantum scattering calculations to heavy diatomic molecules and to polyatomic molecules with a high density of states. 
For chemical reactions at elevated temperatures, the goal is to extend rigorous calculations of reactive collisions in external fields to the multiple partial wave scattering regime. 
Compared to ultracold $s$-wave scattering, higher temperatures allow for more pathways of angular momentum coupling, opening new reaction channels, while also increasing the complexity of the scattering calculations.  
Both of these goals are important for an emerging research field focused on chemical reactions at temperatures on the order of a few Kelvin \cite{Krems:08,Balakrishnan:16,Bohn:17,Karman:24}.

It is important to point out that significant insight into low-temperature scattering dynamics can be gained through semi-analytical or model approaches focusing on long-range physics \cite{Idziaszek:10,Quemener:12}. For example, approximate universal models (UMs) \cite{Idziaszek:10} suggest that averaged properties of many alkali-dimer chemical reactions, such as the total inelastic loss rates, can be understood in terms of the formation of an absorbing short-range reaction complex \cite{Mayle:12,Mayle:13}. These models can be used to develop strategies to shield utracold molecules from chemical losses by external fields inducing long-range barriers \cite{Gorshkov:08,Karman:18,Lassabliere:18}. 
However, by design, UMs focus on long-range interactions and rely on the capture (or absorbing) boundary condition at short range. Because it is  impossible to define such a  boundary condition in a rigorous quantum mechanical way  \cite{Rackham:01}, the Wentzel-Kramers-Brillouin (WKB) approximation is often used,  which is equivalent to applying an imaginary absorbing potential 
 in the short-range reaction complex region. This  procedure 
results in a non-unitary scattering matrix \cite{Rackham:01,Rackham:03a} that cannot provide rigorous information about state-to-state reaction observables and product state distributions, which can now be measured for ultracold reactions using full coincidence detection \cite{Liu:22}. 
UMs do not explain large deviations from the universal behavior recently observed  in  ultracold K~+~NaK, Na~+~NaLi, and NaLi~+~NaLi collisions near magnetic Feshbach resonances  \cite{Yang:19,Wang:21,Park:23,Park:23b,Karman:23} or the sensitivity of K~+~NaK  collisions to the choice of the initial hyperfine states of NaK and K \cite{Voges:22}. 
Clear signatures of  non-universal behavior have recently been observed in rigorous quantum scattering calculations on the ultracold chemical reaction Li($^2$S)~+~NaLi$(a^3\Sigma^+)$ $\to$ Na($^2$S)~+~Li$_2 (a^3\Sigma_u^+)$,  suggesting the  possibility of refining {\it ab initio} interaction PESs based on the experimental measurements of  total reaction rates and product state distributions \cite{Morita:23}, as  recently  demonstrated in a series of precision  experimental and theoretical studies of Feshbach resonances
in cold  He~+~H$_2^+$ and Ne~+~H$_2^+$ collisions \cite{Horn:24,Horn:25}. Leveraging such possibilities  requires numerically exact, fully converged  quantum reactive scattering calculations including  interactions with  external electromagnetic fields commonly present in most experiments.


While converged quantum scattering calculations for atom - molecule inelastic scattering in electric and magnetic fields 
can now be performed for realistic, deeply anisotropic interaction PESs \cite{Tscherbul:11,Morita:18,Morita:20,Morita:24}, rigorous theory of chemically reactive scattering in external fields remains a significant challenge \cite{Tscherbul:08,Tscherbul:15}. 
Quantum reactive scattering calculations are computationally demanding even in the absence of external fields
due to multiple reaction arrangements and a large number of ro-vibrational states participating in the reaction dynamics \cite{Althorpe:03}. 
The presence of external fields breaks the spherical symmetry and further
complicates the numerical calculations by introducing couplings between states of different total angular momenta of the collision complex. 
Within the time-independent quantum scattering formalism, this leads to a large number of coupled differential equations to be solved in the coordinate representation that account simultaneously for multiple chemical reaction arrangements and multiple total angular momentum states.  

Previously, we demonstrated that converged reaction probabilities for ultracold atom-molecule chemical reactions in moderate electric fields can be obtained \cite{Tscherbul:15} by using a total angular momentum (TAM) basis set. This was accomplished by  simultaneously propagating
 multiple field-coupled TAM states into the asymptotic region, where the quantum states of the reaction complex can be projected onto the quantum states of collision partners with well-defined chemical identity. This asymptotic projection, however, involves multi-dimensional integrals that present a significant computational challenge. Here, we develop a numerically efficient frame transformation between the hyperspherical coordinates accounting for multiple chemical arrangements and Jacobi coordinates preserving the chemical identity of reactants and products in the presence of an external electric field.  This allows us to extend reactive scattering calculations for the chemical reactions LiF$(v=1,j=0)$~+~H $\to$ Li~+~HF and F + HD$(v=0,j=0)$ $\to$ HF + D, DF~+~H to a wider range of collision energies and angular momenta than was previously feasible. 

In this article, we use this more efficient formalism to examine the response of chemical reactions to electric fields both in the regime of $s$-wave reactive scattering and in the multiple partial wave scattering regime. 
 For the chemical reaction of highly polar LiF molecules  with H atoms at a collision energy of 14 mKelvin, we observe resonant structure in the field dependence of the reactive scattering cross sections. This structure can be attributed to tunneling-driven interactions between different chemical arrangements. We also observe that an external electric field can have a significant effect on ultracold reactive scattering, even in the absence of scattering resonances, due to field-induced couplings between states of different total angular momenta. Our analysis demonstrates that rigorous calculations of cross sections for reactions of polar molecules in electric fields require multiple total angular momentum states for convergence, even for $s$-wave reactions of molecules in the ground rotational state $j=0$.
 This indicates that high-order field-induced couplings have a significant effect on the reactive scattering cross sections, making ultracold reactions of polar molecules effectively engage non-zero partial wave states. For the archetypal F~+~HD $\to$ HF~+~D, DF~+~H reaction at 1 Kelvin, we observe  no significant field effects on either the total reaction cross section or rovibrationally state-resolved product state distributions at 200 kV/cm, the largest dc electric field readily attainable in the laboratory \cite{Krems:08}. We also find that  calculations with restricted basis sets produce significant spurious field effects on product-state distributions. These effects vanish as the quantum scattering calculations become fully converged. This illustrates the importance of rigorous calculations and basis set convergence for reliable conclusions regarding the effects of external fields on chemical reactivity. 
Our calculations provide the first rigorous test of the effects of electric fields on a chemical reaction of polar molecules at collision energies near 1 Kelvin.



\section{Quantum theory of reactive scattering in dc electric fields}
\label{sec:theory}


The present work is based on the rigorous formalism for quantum reactive scattering using hyperspherical coordinates, originally developed by Schatz \cite{Schatz:88} and more recently implemented in the ABC computer program \cite{Skouteris:00}. Because this approach is well documented elsewhere, we describe only the necessary details, focusing instead on the new methodological developments pertaining to the interaction of the reaction complex with external electric fields.  The present approach extends our original formulation of quantum reactive scattering in external fields~\cite{Tscherbul:08} by recasting the method in the total angular momentum (TAM) representation and the work in Ref. \cite{Tscherbul:15} by improving the efficiency of the projections of the numerical solutions for chemically reactive complexes to the proper boundary conditions. For the present work,   we have developed an extended version of the quantum reactive scattering program ABC \cite{Skouteris:00} that is capable of performing reactive scattering calculations in the presence of an external electric field and that is much faster than the computer program used in Ref.~\citenum{Tscherbul:15}. (Note that  the uncoupled basis set  approach of Ref.~\citenum{Tscherbul:08} was by far less efficient than that developed in Ref.~\citenum{Tscherbul:15} and was unable to provide converged results). 
 The key new features include: (i) the use of the TAM basis sets with running values of total angular momentum $J$ and parity $\eta$, and (ii)  the modified reactive scattering boundary conditions as described in Sec.~\ref{sec:BC} below.

\subsection{Hamiltonian in hyperspherical coordinates}
\label{sec:Hamiltonian}

We consider an atom-diatom chemical reaction with three  distinct reaction arrangements, (e.g., Li + HF, H + LiF and F + LiH) and use the Fock-Delves (FD) hyperspherical coordinates: the hyperradius $\rho=\sqrt{R_\alpha^2 + r_\alpha^2}$ and the hyperangles $\theta_\alpha$ and $\gamma_\alpha$ 
 defined as $\tan\theta_\alpha={r_\alpha}/{R_\alpha}$, and $\cos \gamma_\alpha=(\mathbf{R}_\alpha\cdot \mathbf{r}_\alpha)/(R_\alpha r_\alpha)$ in terms of  the mass-scaled Jacobi vectors $\mathbf{R}_\alpha$ and $\mathbf{r}_\alpha$  in arrangement $\alpha = 1,2,3$ \cite{Pack:87}. A body-fixed (BF) coordinate frame is used with the quantization axis defined by the vector  $\mathbf{R}_\alpha$.
Expressed in the FD coordinates, the Hamiltonian of the atom-molecule reaction complex in the presence of an external electric field is \cite{Pack:87,Schatz:88,Tscherbul:08,Tscherbul:15}
\begin{equation}\label{H}
\hat{H} = -\frac{1}{2\mu \rho^5}\frac{\partial}{\partial \rho}\rho^5 \frac{\partial}{\partial\rho} + \frac{ (\hat{\mathbf{J}}-\hat{\mathbf{j}}_{\alpha})^2 }{2\mu \rho^2\cos^2\theta_\alpha} + V(\rho,\theta_\alpha,\gamma_\alpha) + \hat{H}_{\text{mol},\alpha}
\end{equation}
Here, $\hat{\mathbf{J}}$ is the total angular momentum of the reaction complex, $\hat{\mathbf{j}}_\alpha$ is the rotational angular momentum of the diatomic molecule in arrangement $\alpha$, and $V(\rho,\theta_\alpha,\gamma_\alpha)$ is the interaction potential energy surface (PES) of the triatomic reaction complex. 
Here, we assume that the chemical reaction occurs on a single adiabatic PES. Electronically non-adiabatic effects are known to play a significant role in, e.g., the F~+~H$_2$  chemical reaction at low temperatures \cite{Tizniti:14};  they can be incorporated in our approach as described in Refs.~\cite{Alexander:00b, Kendrick:18,Kendrick:21}.
The last term of Eq. (\ref{H}) describes the isolated reactants and products placed in an external dc electric field \cite{Tscherbul:08,Tscherbul:15}
\begin{equation}\label{Hmol}
\hat{H}_{\text{mol},\alpha}= \frac{-1}{2\mu \rho^2\sin^22\theta_\alpha}\frac{\partial}{\partial \theta_\alpha}\sin^22\theta_\alpha \frac{\partial}{\partial\theta_\alpha} + \frac{ \hat{\mathbf{j}}_{\alpha}^2 }{2\mu \rho^2\sin^2\theta_\alpha} + V_\alpha (\rho,\theta_\alpha) - \mathbf{d}_\alpha(\rho,\theta_\alpha)\cdot \mathbf{E},
\end{equation}
where $\mathbf{d}_\alpha$ is the electric dipole moment of the diatomic molecule in arrangement $\alpha$ and $\mathbf{E}$ is the applied electric field and $V_\alpha (\rho,\theta_\alpha) $ is the diatomic vibrational potential in arrangement $\alpha$.
The direction of $\mathbf{E}$ defines the space-fixed (SF) quantization axis.

\subsection{Total angular momentum representation}
\label{sec:TotalJBasis}

The wavefunction of the reaction complex is expanded as
\begin{align}\label{BF}
\Psi = {\rho}^{-5/2} \sum_i F_i(\rho) \Phi_i(\rho;\Omega),
\end{align} 
where $\Phi_i(\Omega)$ are  the adiabatic states (also known as adiabatic surface functions \cite{Pack:87})  obtained by solving the adiabatic eigenvalue problem 
\begin{equation}\label{AEVP}
\hat{H}_\text{ad}\Phi_i(\Omega;\rho)=\epsilon_i(\rho)\Phi_i(\Omega;\rho).
\end{equation}
Here, $\epsilon_i(\rho)$ are the adiabatic hyperspherical energies, and $\hat{H}_\text{ad}$ is the adiabatic surface Hamiltonian obtained by subtracting the hyperradial kinetic energy from the full Hamiltonian in Eq. (\ref{H})  \cite{Pack:87,Schatz:88,Tscherbul:08,Tscherbul:15}
\begin{equation}\label{Had}
\hat{H}_\text{ad} = \frac{ (\hat{\mathbf{J}}-\hat{\mathbf{j}}_{\alpha})^2 }{2\mu \rho^2\cos^2\theta_\alpha} + V(\rho,\theta_\alpha,\gamma_\alpha) + \hat{H}_{\text{mol},\alpha}.
\end{equation}
 To solve the eigenvalue problem, we expand the adiabatic states  as \cite{Pack:87,Schatz:88,Tscherbul:08,Tscherbul:15}
\begin{equation}\label{adiabaticExp}
\Phi_i(\rho;\Omega) =  \sum_{J, \eta} \sum_{\alpha,v,j, k} W_{\alpha vjJk\eta,i}  |\alpha v j Jk \eta\rangle
\end{equation}
where the bare (zero-field) FD basis states in the BF/TAM representation \cite{Tscherbul:12}
\begin{equation}\label{primitiveFD}
|\alpha vjJk\eta\rangle = |JMk\eta\rangle \frac{2 \chi_{\alpha vj}(\theta_\alpha;\rho)}{\sin 2\theta_\alpha}
\end{equation}
are composed of  the bare FD ro-vibrational basis functions  $\chi_{\alpha vj}(\theta_\alpha;\rho)$ and  the symmetry-adapted BF angular basis functions $|JMk\eta\rangle$ defined below. The primitive basis \eqref{primitiveFD} is the same as that used in the hyperspherical approach of Schatz \cite{Schatz:88}  implemented in the ABC code. The states \eqref{primitiveFD}  form an overcomplete basis set in the strong interaction region of small $\rho$, which is canonically orthogonalized within each $\rho$ sector  \cite{Skouteris:00}. The expansion coefficients $W_{\alpha vjJk\eta,i}(\rho) $ are the components of the eigenvector matrix, which diagonalizes the adiabatic Hamiltonian at the center of each $\rho$ sector.

The symmetry-adapted BF/TAM basis states in \eqref{primitiveFD}
\begin{equation}\label{symmetric_top_basis}  
|JMk\eta\rangle = N_{k}\left[  |JMk\rangle |jk\rangle + \eta(-1)^{J} |JM-k\rangle |j-k\rangle  \right],
\end{equation}
are composed of the spherical harmonics $|jk\rangle \Rightarrow \sqrt{2\pi}Y_{jk}(\theta_\alpha,0)$ and the symmetric top eigenstates $|JMk\rangle$, where $\eta$ is the inversion parity,  $M$ and $k$  are the projections of $J$ on the SF and BF quantization axes, respectively \cite{Zare:88},  and $N_k=[2(1+\delta_{k0})]^{-1/2}$. We note that the $k=0$ parity-adapted basis states (which correspond, e.g., to  $j=0$  rotational states of the reactants and products) vanish if $ \eta(-1)^{J}= -1$.
Thus, nontrivial $k=0$ parity-adapted basis states  \eqref{symmetric_top_basis}  only exist for $ \eta=(-1)^{J}$.

In the absence of electric fields, $J$ and $\eta$ are good quantum numbers, the summation over $J$ and $\eta$ in Eq.~\eqref{adiabaticExp} can be dropped,   and  the present approach reduces to the well-established theories of Schatz \cite{Schatz:88} and Pack and Parker \cite{Pack:87}.  The interactions of molecules with external electric fields
break the spherical and inversion symmetries of the free space and 
 induce couplings between states of different total angular momentum and $\eta$, making it necessary to consider multiple bocks of the Hamiltonian matrix corresponding to different values of $J$ and $\eta$  simultaneously. 
The matrix elements of the molecule-field interaction, $\hat{H}_{S}= - \mathbf{d}_\alpha(\rho,\theta_\alpha)\cdot \mathbf{E}$ [see Eq.~\eqref{Hmol}] in each reaction arrangement \cite{Tscherbul:15}
\begin{multline}\label{mxelEfield}
\langle \alpha v j Jk\eta | \hat{H}_{S}| \alpha' v' j' J' k'\eta'\rangle =   
-d_\alpha E \langle  \chi_{\alpha vj}(\theta_\alpha;\rho)| \chi_{\alpha v'j'}(\theta_{\alpha};\rho)\rangle  \frac{\delta_{\alpha\alpha'} }{[(1+\delta_{k0})(1+\delta_{k'0})]^{1/2}} \\ \times [(2J+1)(2J'+1)(2j+1)(2j'+1)]^{1/2}  (-1)^M \delta_{\eta+\eta',0} \threejm{J}{M}{1}{0}{J'}{-M} \threejm{j}{0}{1}{0}{j'}{0} \\ \times
\left[ \threejm{J}{k}{1}{k'-k}{J'}{-k'} \threejm{j}{-k}{1}{k-k'}{j'}{k'} + \eta' (-)^{J'} \threejm{J}{k}{1}{-k'-k}{J'}{k'} \threejm{j}{-k}{1}{k+k'}{j'}{-k'}   \right]
\end{multline}
vanish unless $j=j\pm1$, $J=J\pm 1$, and $\eta'=-\eta$, leading to a tridiagonal form of the molecule-field interaction Hamiltonian. 
The  electric field hybridizes bare rotational states of the reactants and products, resulting in pendular (or field-dressed) Stark states \cite{Friedrich:91,Friedrich:91b,Rost:92} in the limit of large $\rho$. These states  transform progressively into those of the reaction complex with decreasing $\rho$.

 It may seem counterintuitive to use the TAM basis (\ref{primitiveFD}) to describe chemical reactions  in the presence of external electric fields.
 However, Eq.~\eqref{mxelEfield} shows that the matrix of the field-induced interaction in the basis (\ref{primitiveFD}) is tridiagonal in $J$ and thus only a limited number of $J$-states is generally required for a fully converged scattering  calculation even in  strong electric fields \cite{Tscherbul:12}. This offers a tremendous computational advantage over the originally proposed approach to reactive scattering in electric fields \cite{Tscherbul:08}, which used a fully uncoupled basis representation and did not take advantage of  the tridiagonal structure of the Hamiltonian matrix.

  To illustrate this advantage, consider a TAM basis with  $j_\text{max}=17$ and $J_\text{max}=3$, which is sufficient to obtain converged results for the LiF~+~H reaction at low electric fields (see below). The basis has  268 functions per vibrational state for the total angular momentum projection $M=0$.
 On the other hand, the number of functions $|jm\rangle |l m_l\rangle$ in the fully uncoupled basis  set of Ref.~\cite{Tscherbul:08} with the comparable cutoff parameters $j_\text{max}=17$ and $l_\text{max}=17$ is 3,894 per one vibrational state, which is 14.5 times larger.
 The computational effort of solving coupled-channel equations scales as ${\cal O}(N^3)$ with the number of basis states, so the TAM basis is $\simeq$3,000 times more computationally efficient already for the modest value of $j_\text{max}$.
 To our knowledge, the largest CC calculation ever performed contained 18,852 channels  \cite{Suleimanov:12}.
  More rotational states are required to describe atom-molecule collisions and chemical reactions involving heavier collision partners and/or more strongly  anisotropic interactions. For example, Rb~+~SrF collisions in an external magnetic field are characterized by a deep and strongly anisotropic PES,   requiring  175 rotational states for convergence \cite{Morita:18,Morita:24}. Neglecting electronic and nuclear spins, this yields  2,796 coupled channels in the TAM basis with $J_\text{max}=3$  to be compared with 3,634,576 channels in the fully uncoupled basis with $j_\text{max}=175$ and $l_\text{max}=175$. While the former case can be easily solved using modest computational resources, the latter is completely intractable as it is currently impossible to even store the 3,634,576-channel coupling matrix  in computer memory. This highlights the tremendous computational potential of the TAM basis, which  has already enabled converged quantum scattering computations on strongly anisotropic atom-molecule collisions in an external magnetic field \cite{Tscherbul:10,Tscherbul:12,Tscherbul:11,Suleimanov:12,Morita:18,Morita:20,Morita:24}. Prior to the development of the TAM basis for collisions in external fields \cite{Tscherbul:10,Tscherbul:12,Suleimanov:12}, these systems were out of reach of rigorous computational methodology. Here, we leverage this potential for chemical reactions in electric fields.
    



We substitute the expansion of $\Psi$ in adiabatic surface functions ~\eqref{BF} into the time-independent Schr{\"o}dinger equation with total energy $E$. This results in a system of adiabatic coupled-channel (ACC) equations \cite{Pack:87,Skouteris:00,Schatz:88,Nyman:00}, which are decoupled  within each hyperradial sector due to Eq.\eqref{AEVP}
\begin{equation}\label{ACC}
\biggl{[} \frac{d^2}{d\rho^2} - \frac{15}{8\mu \rho^2} + 2\mu[E-\epsilon_i(\rho)]\biggr{]}F_i(\rho) = 0.
\end{equation}
The ACC equations are integrated numerically  on a grid of hyperradial sectors $\rho_i$ by propagating the log-derivative matrix across each sector and then transforming it to the locally adiabatic basis of the next sector, as described in detail, e.g., in Refs.~\citenum{Pack:87,Tscherbul:08}. This includes diagonalizing the adiabatic Hamiltonian matrix at the middle of each  hyperradial sector and then using the resulting adiabatic potentials $\epsilon_i(\rho)$ and eigenvectors $W_{ji}(\rho)$ to advance the log-derivative matrix from to the next sector.


Upon reaching the asymptotic region of large $\rho=\rho_{a}$, where the couplings between different reaction arrangements (as well as between different rovibrational states within the same arrangement) can be neglected \cite{Skouteris:00},   the log-derivative matrix  is transformed from the adiabatic basis to the primitive FD basis.
In order to extract  the reaction probabilities from this log-derivative matrix, it is necessary to transform it from the hyperspherical coordinates to Jacobi coordinates, 
in which the scattering boundary conditions are applied \cite{Pack:87}.  This
 involves evaluating the wavefunction of the reaction complex, along with its derivatives, at the boundary of a region defined in hyperspherical coordinates, reprojecting it to the corresponding region in Jacobi coordinates, and finally matching to the asymptotic solutions in Jacobi coordinates expressed in terms of  the Riccati-Bessel functions \cite{Johnson:73,Pack:87}. These transformations, which we refer to as ``wavefunction surgery'', are well-documented for atom-diatom chemical reactions in the absence of external fields \cite{Pack:87}.  Below we describe an efficient extension of these transformations to chemical reactions in the presence of external  electric fields.


\subsection{Wavefunction surgery for reactive scattering boundary conditions in electric fields}
\label{sec:BC}


The purpose of rigorous quantum scattering calculations is to compute the full $S$ matrix encoding the amplitudes for probabilities of elastic and inelastic transitions as well as chemical transformations. The $S$ matrix is obtained by matching the numerically computed log-derivative matrix to the quantum scattering boundary conditions. 
These boundary conditions are most naturally applied in the Jacobi coordinates \cite{Pack:87}.
  External fields induce couplings between molecular states within the reactant and product arrangements, making the asymptotic Hamiltonians \eqref{Hmol} non-diagonal in the basis of field-free molecular states. Therefore, boundary conditions cannot be applied in the basis of field-free states and the asymptotic log-derivative matrix must be transformed to the basis of field-dressed states  \cite{Krems:04}, before the $S$ matrix is constructed from the log-derivative matrix. The original asymptotic projection procedure employed for reactive scattering calculations in the absence of fields \cite{Pack:87} must therefore be modified to include this additional transformation, as described below.



With $\rho=\rho_{a}$ in the asymptotic region, the different reaction arrangements are completely decoupled and the boundary conditions on the scattering wavefunction can be expressed in Jacobi coordinates \cite{Pack:87} suitably generalized to include the modification of the asymptotic states of the reactants and products by external fields \cite{Krems:04,Tscherbul:12}
 \begin{equation}\label{Jac_expansion}
|\Psi^{i} \rangle_\text{Jac} = \sum_{m} \frac{1}{R_{\alpha_m} r_{\alpha_m}}  F^{i}_{m} (R_{\alpha_m})  |m \rangle_\text{Jac},
\end{equation}
where the index ``$i$'' refers to the initial state of the reactants prior to reaction,  and the field-dressed  basis states in Jacobi coordinates $|m \rangle_\text{Jac}= |\alpha_m v_m \gamma_m l_m \rangle_\text{Jac}  $ are given in terms of field-free Jacobi basis states  $|m'\rangle_\text{Jac} = |\alpha_{m'}  v_{m'} J_{m'} j_{m'} \gamma_{m'} l_{m'} \rangle_\text{Jac}$ as
\begin{align}\label{m_Jac}
|m \rangle_\text{Jac} &=  \sum_{m'} C^\text{Jac}_{m' m}(E) |m'\rangle_\text{Jac}\\ \label{m_Jac_prime}
|m'\rangle_\text{Jac} &= \xi_{\alpha_{m'} v_{m'} j_{m'}} (r_{\alpha_{m'}})  \mathcal{J}^{J_{m'} M}_{j_{m'} l_{m'}} (\hat{R}_{\alpha_{m'}};\hat{r}_{\alpha_{m'}}).
\end{align}
Here,  
$ \xi_{\alpha_{m'} v_{m'} j_{m'}} (r_{\alpha_{m'}})$ is the rovibrational eigenfunction of the molecule in arrangement $\alpha_{m'}$ with vibrational and rotational quantum numbers $v_{m'}$ and $j_{m'}$, $C^\text{Jac}_{m' m}(E) $ are the Stark mixing coefficients
computed by diagonalizing the asymptotic Hamiltonian $\hat{H}_\text{as}$ in the  bare Jacobi basis \eqref{m_Jac_prime}.
Because the asymptotic Hamiltonian 
does not contain couplings between the basis states of different chemical arrangements and orbital angular momenta, its eigenvector matrix  $\mathbf{C}^\text{Jac}=\{C_{m'm}^\text{Jac} \}$ is diagonal in $\alpha$ and $l$.
Throughout this section, we label field-dressed basis functions by unprimed indices collectively representing their quantum numbers, {e.g.},  $n = \{ \alpha_n, v_n, \gamma_n, l_n \}$ (FD coordinates) and $m = \{ \alpha_m, v_m, \gamma_m, l_m \}$ (Jacobi coordinates).
The primed indices are reserved for bare basis functions, {e.g.}, $n' = \{\alpha_{n'}, v_{n'}, J_{n'}, j_{n'}, l_{n'},J_{n'}\}$ (FD coordinates) and  $m' = \{\alpha_{m'}, v_{m'}, J_{m'}, j_{m'}, l_{m'}, J_{m'}\}$ (Jacobi coordinates). The four types of basis functions used in this section are summarized in Table I. 

%
\begin{table}
 \caption{\label{} Types of basis functions used in quantum reactive scattering calculations in the presence of external electric fields. The subscripts ``Jac'' (``FD'') are sometimes added to the state labels $|m\rangle$, $|m'\rangle$ ($|n\rangle$, $|n'\rangle$) and eigenvector components for clarity. }
 \begin{tabular}{ccc}
 \hline
 \hline
  Basis function  & Notation & Equation \\
  \hline
 Jacobi, field-free  & $|m'\rangle = \xi_{\alpha_{m'} v_{m'} j_{m'}} (r_{\alpha_{m'}})  \mathcal{J}^{J_{m'} M}_{j_{m'} l_{m'}} (\hat{R}_{\alpha_{m'}};\hat{r}_{\alpha_{m'}})$ & (\ref{m_Jac}) \\
  Jacobi, field-dressed & $|m \rangle =  \sum_{m'} C^\text{Jac}_{m' m}(E) |m'\rangle$ &  \eqref{m_Jac_prime} \\
    Fock-Delves, field-free & $|n'\rangle =  \left[ {2 \chi_{\alpha_{n'} v_{n'}j_{n'}}(\theta_{\alpha_{n'}};\rho)} / {\sin 2\theta_{\alpha_{n'}}}  \right]  \mathcal{J}^{J_{n'}M}_{j_{n'}l_{n'}} (\hat{R}_{\alpha_{n'}};\hat{r}_{\alpha_{n'}})$ & \eqref{FDbasis0field} \\
      Fock-Delves, field-dressed & $|n\rangle = \sum_{n'} C^\text{FD}_{n'n}  |n'\rangle$ & \eqref{n_FD} \\
       \hline
 \hline
 \end{tabular}
 \end{table}

  The eigenfunctions of the total angular momentum (TAM) of the reaction complex in the space-fixed (SF) coordinate frame, also known as bipolar spherical harmonics \cite{Pack:87,Zare:88,VarshalovichBook}, are given by
 \begin{equation}\label{FDbasisBipolar}
\mathcal{J}^{JM}_{jl} (\hat{R}_{\alpha};\hat{r}_{\alpha}) = \sum_{m_j, \,m_l} (-1)^{j-l+M} (2J+1)^{1/2}\threejm{j}{m_j}{l}{m_l}{J}{-M} Y_{jm_j} (\hat{r}_\alpha) Y_{lm_l}(\hat{R}_\alpha)
\end{equation}
where  $(:::)$ denote 3-$j$ symbols and the $m'$ subscripts have been ommitted for brevity. They are obtained by vector coupling of the eigenstates $Y_{jm_j} (\hat{r}_\alpha)$ of $\hat{N}^2$ and $\hat{N}_z$  and eigenstates $Y_{lm_l}(\hat{R}_\alpha)$ of $\hat{L}^2$ and $\hat{L}_z$. The $z$-axis of the SF frame is assumed to lie along the direction of the external electric field, so that the total angular momentum projection $M$ is a good quantum number \cite{Krems:04,Tscherbul:06,Abrahamsson:07}.

The asymptotic form of the Jacobi radial solutions \eqref{Jac_expansion} in each reaction arrangement can be written in terms of the reactance (K)  matrix elements  \cite{Johnson:73,Pack:87} as follows:
\begin{equation}\label{BoundaryConditions}
F_{mi}(R_{\alpha_{m'}} \to \infty) \simeq \delta_{mi} a_{mm}(R_{\alpha_{m'}}) - b_{mm}(R_{\alpha_{m'}}) K_{mi}, 
\end{equation}
where $K_{ni}$ are the $K$-matrix elements and the functions $a_n$ and $b_n$ are proportional to the Riccati-Bessel functions or modified Bessel functions of the third kind depending on whether the asymptotic scattering channel $n$ is open or closed \cite{Johnson:73,Pack:87}.  The $S$-matrix elements can be computed from the $K$-matrix elements, as usual \cite{Johnson:73,Pack:87}. 
However, instead of the Jacobi radial solutions  needed in  Eq.~\eqref{BoundaryConditions}, the ACC calculations provide  {\it hyperradial solutions in the FD coordinates}.
 In order to transform the hyperradial solutions to Jacobi coordinates, we generalize  the procedure of Pack and Parker \cite{Pack:87} to include the effects of external electric fields.

 
 Our goal is to derive the asymptotic transformation between the radial solutions in Jacobi coordinates $F_n^i(R)$ defined by Eq.~\eqref{Jac_expansion} and the hyperradial solutions obtained by numerical propagation of the adiabatic CC equations \eqref{ACC}. 
 The ``raw'' log-derivative matrix available by propagating the adiabatic CC equations is expressed in the BF adiabatic basis  \eqref{adiabaticExp}.
We first back-transform the matrix using the adiabatic eigenvector matrix evaluated at $\rho=\rho_a$ to obtain the log-derivative matrix in the field-free BF/TAM FD basis \eqref{primitiveFD}. However, because the asymptotic Hamiltonian in this latter basis is not diagonal, we need two additional transformations, first to the  field-free SF/TAM FD basis, and then  to the field-dressed  SF/TAM  FD basis (the last line in Table I). Applying these transformations   gives
$\mathbf{Y}(\rho_a)=\bm{\Gamma}'(\rho_a)[\bm{\Gamma}(\rho_a)]^{-1}$ with 
  the square $N\times N$ matrices of hyperradial solutions  $\bm{\Gamma}(\rho)=\{\Gamma^{i}_{n} (\rho)\} =\{\Gamma_{ni}(\rho)\}$ defined as
 \begin{equation}\label{FDexpansion}
|\Psi^{i} \rangle_\text{FD} = \rho^{-5/2} \sum_{n} \Gamma^{i}_{n} (\rho) |n \rangle_\text{FD}.
\end{equation}
Here, $|n \rangle_\text{FD}$ are  field-dressed hyperangular basis functions  in the SF frame
 \begin{equation}\label{n_FD}
|n\rangle_\text{FD} = \sum_{n'} C^\text{FD}_{n'n}  |n'\rangle_\text{FD},
\end{equation}
and their zero-field counterparts $|n'\rangle$ are given by
 \begin{equation}\label{FDbasis0field}
|n'\rangle =  \left[ \frac{2 \chi_{\alpha_{n'} v_{n'}j_{n'}}(\theta_{\alpha_{n'}};\rho)}{\sin 2\theta_{\alpha_{n'}}}  \right]  \mathcal{J}^{J_{n'}M}_{j_{n'}l_{n'}} (\hat{R}_{\alpha_{n'}};\hat{r}_{\alpha_{n'}}).
\end{equation}
 The dependence of these basis states on the orientation of the Jacobi vectors $\hat{R}_{\alpha_{n'}}$ and $\hat{r}_{\alpha_{n'}}$ is given by the same bipolar spherical harmonics  as in the FD case \eqref{FDbasisBipolar}, so the only difference between Eqs.~\eqref{m_Jac_prime} and \eqref{FDbasis0field}  is in the vibrational basis functions: $ {2 \chi_{\alpha_{n'} v_{n'}j_{n'}}(\theta_{\alpha_{n'}};\rho)}/({\sin 2\theta_{\alpha_{n'}}}) $  (FD) vs.  $\xi_{\alpha_{n'} v_{n'} j_{n'}} (r_{\alpha_{n'}})$ (Jacobi).
 This similarity leads to a significant simplification of the  transformation between the two basis sets.
The coefficients $C^\text{FD}_{n'f}$ in Eq.~\eqref{n_FD} describe electric field-induced Stark mixing within the reactant and product arrangements in the FD basis, and are obtained by diagonalizing the asymptotic Hamiltonian in the zero-field FD basis of Eq.~\eqref{FDbasis0field}. Note that  these coefficients are distinct from those in Eq.~\eqref{m_Jac}.

To perform the coordinate transformation, we first use   
 the orthonormality of field-dressed FD basis functions (\ref{n_FD})  at a sufficiently large value of $\rho=\rho_a$ to obtain
 \begin{equation}\label{starting_point}
 \Gamma_{ni}(\rho)  = \langle n | \rho^{5/2} \Psi^{i}  \rangle_\text{FD},
 \end{equation}
 where $\langle \ldots \rangle_\text{FD}$ indicates integration over the angular variables $\hat{R}_\alpha$ and $\hat{r}_\alpha$  as well as the FD hyperangle $\theta_\alpha$.
  We next substitute  $|\Psi^{i}  \rangle_\text{Jac}$ from Eq. (\ref{Jac_expansion}) and use the decoupling between the different arrangements at large $\rho=\rho_a$ and the orthonormality of bipolar spherical harmonics within each arrangement to obtain (see Appendix A)
  \begin{equation}\label{Gamma_ni}
\Gamma_{ni} (\rho) =  \sum_m \int  X_{nm}(\theta_{\alpha_{m}},r_{\alpha_{m}};\rho) F_{mi}(R_{\alpha_{m}})d\theta_{\alpha_{m}},
\end{equation}
where the elements of the hyperangle-dependent transformation matrix $\mathbf{X}(\theta_{\alpha_{m}},r_{\alpha_{m}};\rho) $ in the field-dressed basis 
  \begin{equation}\label{X_nm}
X_{nm}(\theta_{\alpha_{m}},r_{\alpha_{m}};\rho)  = \sum_{n',m'}C_{n'n}^\text{FD}(E)     X^{(0)}_{n'm'}(\theta_{\alpha_{m'}},r_{\alpha_{m'}};\rho) C_{m'm}^\text{Jac}(E),
\end{equation}
  are obtained by transforming a product matrix composed of the field-free FD and Jacobi vibrational functions 
   \begin{equation}\label{X_npmp}
   X^{(0)}_{n'm'}(\theta_{\alpha_{m'}},r_{\alpha_{m'}};\rho) = \rho^{1/2} 
   \delta_{\alpha_{n'}\alpha_{m'}} \delta_{J_{n'} J_{m'}}\ \delta_{j_{n'} j_{m'}}\delta_{l_{n'} l_{m'}}
 \chi_{\alpha_{n'}v_{n'}j_{n'}}(\theta_{\alpha_{n'}};\rho)   \xi_{\alpha_{m'} v_{m'} j_{m'}} (r_{\alpha_{m'}}) 
  \end{equation}
  to the field-dressed basis using the Stark mixing coefficients $C_{n'n}^\text{FD}(E) $ and $C_{n'n}^\text{Jac}(E)$ defined above.  The product matrix \eqref{X_npmp}  is diagonal in all but vibrational quantum numbers, which reflects the similarity of the FD and Jacobi basis functions. Note that  only primed  indices occur in Eq.~\eqref{X_npmp} as it involves only field-free basis functions.  Similarly, only non-primed indices occur in Eq.~\eqref{Gamma_ni} as it deals with field-dressed functions.
  
  Equation~\eqref{Gamma_ni} establishes a relationship between the asymptotic hyperradial solutions  $\Gamma_{ni}(\rho_a)$ and their Jacobi counterparts. The latter are, in turn, related to the $K$-matrix elements via the boundary condition \eqref{BoundaryConditions}.  Plugging Eq.~\eqref{BoundaryConditions} into Eq.~\eqref{Gamma_ni} gives
    \begin{equation}\label{Gamma_ni_K}
\Gamma_{ni} (\rho) = 
\int d\theta_{\alpha_{i}}  X_{ni}(\theta_{\alpha_{i}},r_{\alpha_{i}};\rho) a_{ii}(R_{\alpha_{i}})
- \sum_m K_{mi}  \int d\theta_{\alpha_{i}}  X_{nm}(\theta_{\alpha_{m}},r_{\alpha_{m}};\rho) b_{mm}(R_{\alpha_{m}}),
\end{equation}
where the asymptotic solutions can be expressed in terms of the Riccatti-Bessel functions $a_{ii}(R_{\alpha_{i}})$ and $b_{mm}(R_{\alpha_{m}})$ for asymptotically open channels ($k^2_n>0$) or modified Riccatti-Bessel functions for closed channels ($k^2_n<0$)  \cite{Pack:87}. Here, $k^2_n=2\mu (E-\epsilon_n) =2\mu E_C$ is the squared wavevector in the asymptotic channel $|n\rangle$ with threshold energy $\epsilon_n$,  $E$ is the total energy, and $E_C$ is the collision energy. Note that all the terms in the integrands of Eq.~\eqref{Gamma_ni_K} 
 depend on  $\theta_{\alpha_m}$ in a non-trivial way  because the mass-weighted Jacobi coordinates are related to the FD coordinates via $R_\alpha=\rho\cos\theta_\alpha$ and $r_\alpha=\rho\sin\theta_\alpha$.
 Unlike Eqs. (116) and (117) of Ref. \cite{Pack:87}, the radial functions $a_n$ and $b_n$ in Eq.~\eqref{Gamma_ni_K} are expressed in the field-dressed Jacobi basis, i.e., the wavevectors $k_n$ entering the arguments of the functions $a_n$ and $b_n$ correspond to the states of the reactants and products in the presence of an electric field.

 
It is instructive to rewrite Eq.~\eqref{Gamma_ni_K}  in matrix form
 \begin{equation}\label{Gamma_ni_Kmx}
\bm{\Gamma} (\rho) = \mathbf{A}(\rho) - \mathbf{B}(\rho)\mathbf{K},
\end{equation}
where $\bm{\Gamma}$, $\mathbf{A}(\rho)$, $\mathbf{B}(\rho)$, and $\mathbf{K}$ are square $N\times N$ matrices, and $N=N_o+N_c$ is the  total number of scattering channels (open, $N_o$, as well as closed, $N_c$). The transformation matrices  $\mathbf{A}(\rho)$ and $\mathbf{B}(\rho)$ are composed of overlap integrals between the vibrational basis functions in Jacobi and FD coordinates evaluated at a fixed value of  $\rho$ in the asymptotic region
 \begin{align}\notag
{A}_{ni} (\rho) &= \int d\theta_{\alpha_{i}}  X_{ni}(\theta_{\alpha_{i}},r_{\alpha_{i}};\rho) a_{ii}(R_{\alpha_{i}}), \\ \label{A_and_B}
{B}_{nm} (\rho) &= \int d\theta_{\alpha_{m}}  X_{nm}(\theta_{\alpha_{m}},r_{\alpha_{m}};\rho) b_{mm}(R_{\alpha_{i}}).
\end{align}
These matrices can be readily computed by one-dimensional numerical quadrature from the known ingredients (see below).
In principle, Eqs.~\eqref{Gamma_ni_K}-\eqref{Gamma_ni_Kmx} can be used to  determine the $K$-matrix from the  hyperradial solution matrix $\bm{\Gamma} (\rho)$. In practice, however, attempts to compute $\bm{\Gamma} (\rho)$ via direct numerical integration of the ACC equations \eqref{ACC} 
suffer from numerical instabilities in the presence of closed channels, whose rapidly growing components destroy the linear independence  of the columns of  $\bm{\Gamma} (\rho)$  \cite{Manolopoulos:86}.  A well-established  strategy to avoid these instabilities \cite{Johnson:73,Manolopoulos:86} is  to propagate the $N\times N$ log-derivative  matrix
$\mathbf{Y}(\rho)=\bm{\Gamma}'(\rho)[\bm{\Gamma}(\rho)]^{-1}$, which is the quantity produced by the numerical integration of the ACC equations using the log-derivative algorithm \cite{Manolopoulos:86,Skouteris:00}.

To extract the $K$-matrix from the log-derivative matrix, an additional step is necessary, which involves the evaluation of the derivative matrix $\bm{\Gamma}'(\rho)$. Taking the derivative of Eq.~\eqref{Gamma_ni_K} with respect to $\rho$, we  obtain
\begin{equation}\label{matrix_dGamma}
\frac{\partial \mathbf{\Gamma (\rho)}}{\partial \rho} = \frac{1}{2\rho} \mathbf{\Gamma}(\rho) + \left[\mathbf{G}(\rho) - \mathbf{H}(\rho)\mathbf{K}\right],
\end{equation}
with the $N\times N$ matrices $\mathbf{G}(\rho)$ and $\mathbf{H}(\rho)$  given by
\begin{align}\label{G_and_H}
{G}_{ni}(\rho) &= \rho^{1/2}    \int d\theta_{\alpha_n}
  \biggl{[}  \cos\theta_{\alpha_n}  \frac{\partial a_{ii}(R_{\alpha_n})}{\partial R_{\alpha_n}} \tilde{X}_{ni}(\theta_{\alpha_{i}},r_{\alpha_{i}};\rho) 
  +
 \sin\theta_{\alpha_n} a_{ii}(R_{\alpha_n})  \frac{\partial \tilde{X}_{ni}(\theta_{\alpha_{i}},r_{\alpha_{i}};\rho) }{\partial r_{\alpha_n}}   
    \biggr{]},  \\ \notag
{H}_{nm}(\rho) &= \rho^{1/2}    \int d\theta_{\alpha_n}
  \biggl{[}  \cos\theta_{\alpha_n}  \frac{\partial b_{mm}(R_{\alpha_n})}{\partial R_{\alpha_n}} \tilde{X}_{nm}(\theta_{\alpha_{n}},r_{\alpha_{i}};\rho) 
  +
 \sin\theta_{\alpha_n} b_{mm}(R_{\alpha_n})  \frac{\partial \tilde{X}_{ni}(\theta_{\alpha_{i}},r_{\alpha_{n}};\rho) }{\partial r_{\alpha_n}}   
    \biggr{]},  
\end{align}
where 
  \begin{equation}\label{X_nm_scaled}
\tilde{X}_{nm}(\theta_{\alpha_{m}},r_{\alpha_{m}};\rho) =  \frac{1}{\rho^{1/2}} {X}_{nm}(\theta_{\alpha_{m}},r_{\alpha_{m}};\rho)
\end{equation}
is a scaled  hyperangle-dependent transformation matrix, which depends on $\rho$ only parametrically [unlike its unscaled counterpart in Eq.~\eqref{X_nm}], facilitating the evaluation of the $\rho$ derivatives via the chain rule (see Appendix A  for details).


Combining the expressions for $\bm{\Gamma}(\rho)=  \mathbf{A}(\rho) - \mathbf{B}(\rho)\mathbf{K}$ \eqref{Gamma_ni_Kmx} and its hyperradial derivative \eqref{matrix_dGamma},  the FD log-derivative matrix takes the form
\begin{equation}\label{LDmatrix}
\mathbf{Y}(\rho) = \left(  \frac{1}{2\rho} \left[\mathbf{A}(\rho) - \mathbf{B}(\rho)\mathbf{K}\right] + \left[\mathbf{G}(\rho) - \mathbf{H}(\rho)\mathbf{K}\right] \right)
 \left[\mathbf{A}(\rho) - \mathbf{B}(\rho)\mathbf{K}\right]^{-1}.
\end{equation}
Solving for $\mathbf{K}$, we  obtain the final expression for the reactance matrix 
\begin{equation}\label{Kmx}
\mathbf{K} = \left[ \left(\mathbf{Y}(\rho) -\frac{1}{2\rho}\mathbf{I}\right) \mathbf{B} - \mathbf{H} \right ]^{-1} 
 \left[ \left(\mathbf{Y}(\rho) -\frac{1}{2\rho}\mathbf{I}\right)  \mathbf{A} -  \mathbf{G}   \right],
\end{equation}
where $\mathbf{I}$ is the unit matrix.
Note that all the matrices which occur in this equation are square $N\times N$ matrices. This is a significant simplification over our previous work \cite{Tscherbul:15}, in which  the $K$-matrix was expressed via three and four-rank tensors, whose evaluation was extremely computationally intensive. The simplification is achieved here by {\it first} transforming the relevant $\theta_\alpha$-dependent matrices to the field-dressed basis via Eq.~\eqref{X_nm}, and {\it then} taking the hyperangular integral in Eq.~\eqref{Gamma_ni_K}. By contrast, in the previous work \cite{Tscherbul:15}, the hyperangular overlap integrals were evaluated first and then transformed to  the field-dressed basis, a procedure whose computational efficiency is severely limited by the $n$ dependence of the integrals (see the Supplemental Material of Ref.  \citenum{Tscherbul:15}).




\color{black}


The $K$-matrix is computed using Eq. (\ref{Kmx}), and then converted to the $S$-matrix as $\mathbf{S}=(\mathbf{I}+i\mathbf{K}^{\rm oo})(\mathbf{I}-i\mathbf{K}^{\rm oo})^{-1}$, where $\mathbf{K}^{\rm oo}$ is the open-open block of the $K$-matrix and $\mathbf{I}$ is the unit matrix \cite{Pack:87}.
 The reaction cross sections are calculated from the $S$-matrix as
 \begin{equation}\label{sigma}
\sigma_{\alpha v \gamma  \to \alpha'v' \gamma' } = \frac{\pi}{k^2_{\alpha v\gamma} }\sum_M \sum_{l,\, l'} P^M_{\alpha v \gamma l  \to \alpha'v' \gamma'l' }
\end{equation}
where 
\begin{align}\label{P}
P^M_{\alpha v \gamma l  \to \alpha'v' \gamma'l' } &= | S^M_{\alpha v \gamma  l \to \alpha'v' \gamma' l'}  |^2   \quad (\alpha\ne \alpha')
\end{align}
is the fully state-resolved reaction probability  and the index $\gamma$ runs over the Stark states of the reactants and products (note that in the zero-field limit, the index $\gamma$ can be replaced with $j$, $J$ and $\eta$).
The field-dressed Stark states of the reactants and products are labeled by the chemical arrangement ($\alpha_m$), vibrational ($v_m$), and Stark ($\gamma_m$) quantum numbers, which label field-dressed Jacobi basis functions $|m\rangle=|\alpha_m v_m \gamma_m l_m\rangle_\text{Jac}$ in Eq.~\eqref{m_Jac}. Note that because each Stark state is a  linear combination of field-free rotational, total angular momentum, and parity basis states described by the quantum numbers $j_{m'}$, $J_{m'}$, and $\eta_{m'}$, these quantum numbers are no longer good in the presence of an electric field.

  \subsection{Computational details}
  
  We have extended the quantum reactive  scattering code ABC \cite{Skouteris:00} to include the interaction of the reaction complex with external electric fields following the theoretical methodology described above.  The following new features have been implemented in the extended ABC (extABC) code: (i)  hyperspherical FD  basis sets with running  $J$ and $\eta$ values, (ii) matrix elements of the Stark Hamiltonian between these FD basis states \eqref{mxelEfield}, and (iii) field-dependent reactive scattering boundary conditions (see Appendix A). The new convergence parameters {\tt  jtotMax\_\,min} and {\tt jtotMax\_\,max} control the number of total angular momentum states in the basis set. The modifications,  are described in detail in the Supplemental Material of Ref.~\citenum{Tscherbul:15}.
  We used the following values of electric dipole moments: 6.33~D for $^7$Li$^{19}$F, 1.83~D for HF, and 1.85~D for DF \cite{Hebert:68,Werner:80}.

   A further modification of the extABC code used here, as compared to our earlier  work \cite{Tscherbul:15} concerns the evaluation of the projection matrices, which are first transformed to the field-dressed basis and then integrated over $\theta_\alpha$ for improved computational efficiency, as described in the previous section.
  Test calculations show that these modifications reduce the computational cost of applying reactive scattering boundary conditions by several orders of magnitude, enabling the use of larger basis set than in our previous work \cite{Tscherbul:15}.
  The computational details specific to the LiF($v=1,j=0$)~+~H  and  F~+~HD chemical reactions, including the convergence parameters, are given in Sec. III.
  

  \section{Results}

\subsection{LiF($v=1,j=0$)~+~H $\to$ HF~+~Li}

\begin{figure}[t!]
	\centering
	\includegraphics[width=0.8\columnwidth, trim = 0 0 0 0]{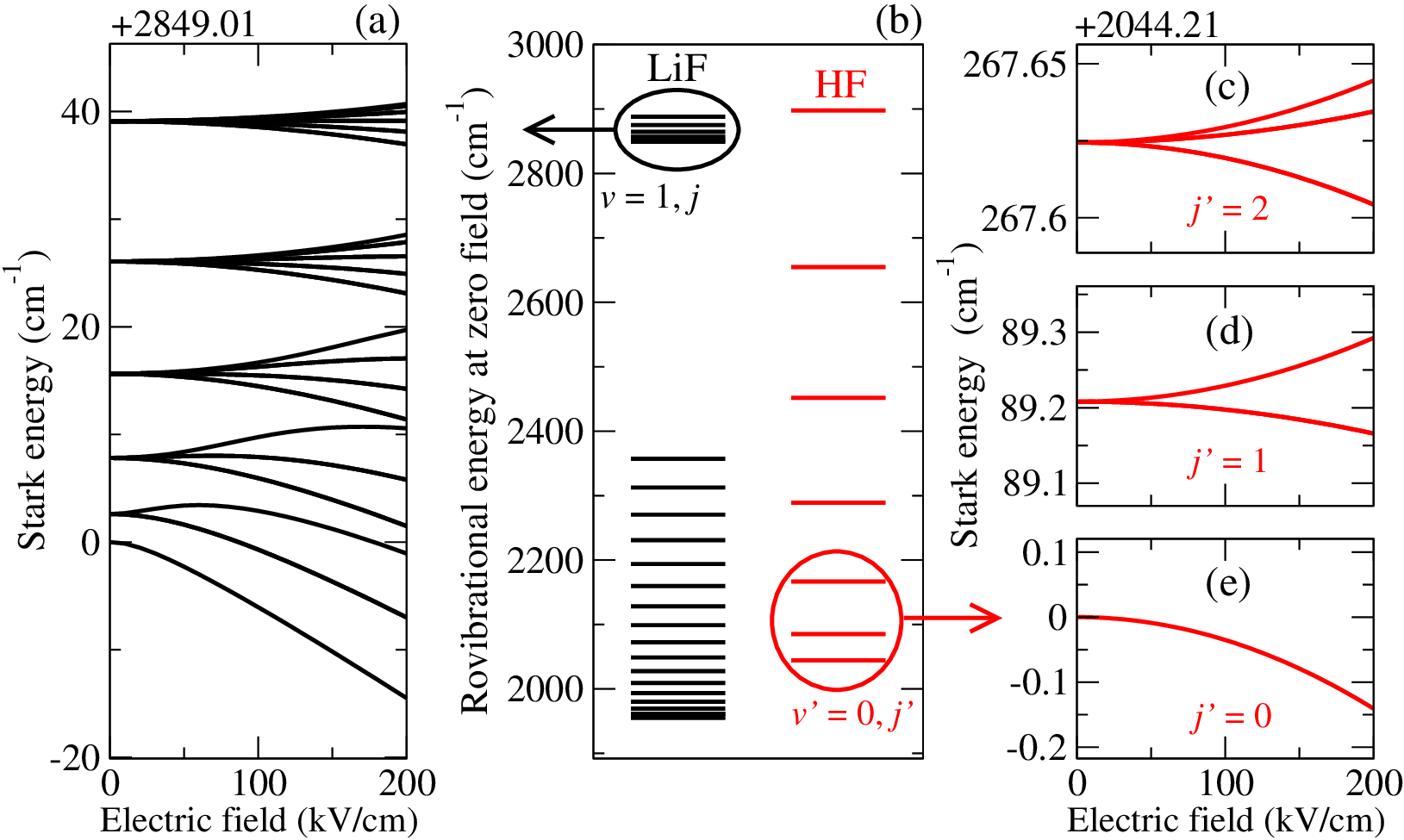}
	\caption{Energetics of the chemical reaction LiF$(v=1,j=0)$~+~H $\to$ Li~+~HF in a dc electric field. The energy is measured from the bottom of the potential well in the reactants valley.  }
	\label{fig:energeticsLiFH}
\end{figure}

We begin by exploring the effects of external electric fields on the quantum dynamics of the ultracold LiF($v=1,j=0$)~+~H $\to$ HF~+~Li chemical reaction. The choice of the reaction is motivated by several considerations: first, the reactant molecule is highly polar (permanent dipole moment  for the molecule in the ground vibrational state is $d=6.3$~D), and thus can be expected to exhibit large electric field-induced orientation effects in the entrance reaction channel \cite{Tscherbul:08}. Second, this reaction as well as its inverse (Li + HF $\to$ LiF + H) has recently been the focus of one experimental \cite{Bobbenkamp:11} and several theoretical  \cite{Weck:05,Hazra:15} studies. In principle, the low-temperature dynamics of the Li + HF  and LiF~+~H chemical reactions can be studied  using an experimental setup consisting of a slow beam of HF (or LiF) molecules colliding with a magnetically trapped  target of Li (or H) atoms \cite{Eichhorn:09}. 


The energy level diagram of the LiF($v=1,j=0$) reaction is displayed in Fig.~\ref{fig:energeticsLiFH}. The chemical reaction of the ground rovibrational state of LiF with   H is endothermic, and hence energetically forbidden at ultralow temperatures.   We therefore choose the $v=1,j=0$ rovibrational state of LiF as our initial state. This provides sufficient energy for the molecule to react with H atoms even in the limit of vanishing collision energy. A total of 6 rotational states of nascent HF($v'=0,j'$) products are open at ultralow collision energies in the Wigner $s$-wave regime. We choose $E_C=0.01$~cm$^{-1}$ for all calculations for this chemical reaction. In the $s$-wave regime, the reaction cross section is dominated by a single value of the total angular momentum projection, $M=0$, and we neglect nonzero $M$  contributions  \cite{Tscherbul:15}.

We perform quantum scattering calculations using the extABC code developed for this work and the same {\it ab initio} PES for the LiHF complex \cite{Aguado:03} (the APW PES) as used in our previous work \cite{Tscherbul:15} and in the field-free calculations of Weck and Balakrishnan \cite{Weck:05} and Hazra and Balakrishnan \cite{Hazra:15}.
The  computation parameters are the same as in Ref.~\cite{Tscherbul:15}. However, in the present work, we employ a larger total angular momentum basis ($J_\text{max}=4$, $N=5850$ channels) to improve the convergence of the reaction cross sections at  electric fields above 40~kV/cm.  Quantum scattering computations with such a basis   were not computationally feasible in Ref.~\citenum{Tscherbul:15} due to a steep scaling of the asymptotic matching procedure with the basis set size.   With the improvements  described in Sec.~II of this work, the $J_\text{max}=4$ calculations are now feasible. A single reactive scattering computation takes approximately 6 days on a Xeon E5-2683 v4 processor  with the clock frequency of 2.1 GHz compared to about 5-6 months  on a similar processor using the matching procedure of   Ref.~\citenum{Tscherbul:15}.

\begin{figure}[t!]
	\centering
	\includegraphics[width=0.75\columnwidth, trim = 0 0 0 0]{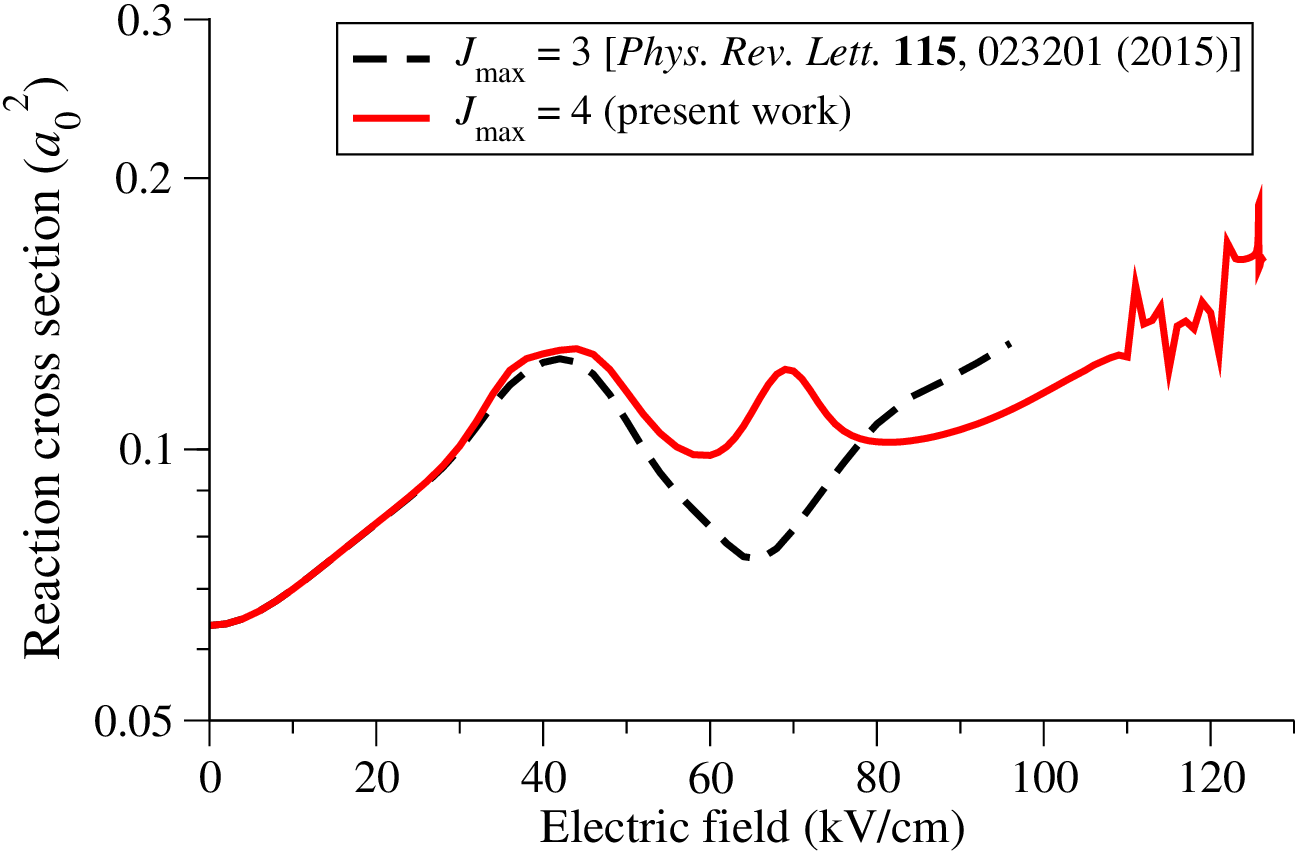}
	\caption{Electric field dependence of the  total integral cross section for the LiF$(v=1,{j}=0)$~+~H $\to$ Li~+~HF($v'=0,{j}'$) chemical reaction summed for all final $j'$ states of the HF$(v'=0)$ product. Full   line -- present calculations. Dashed line -- $J_\text{max}=3$ calculations of Ref.~\citenum{Tscherbul:15}. The collision energy is 0.01 cm$^{-1}$ $\simeq$ 14.4 mKelvin. 
	Note that in the presence of an external electric field, the rotational quantum numbers of the reactants and products $j$ and $j'$  are no longer good quantum numbers, and the pendular quantum numbers  $\gamma$ and $\gamma'$ should be used instead, see Eq.~\eqref{sigma}. Here, $\sigma_{\alpha v j\to \alpha'v'j'}$ denotes the reaction cross section \eqref{sigma} summed over all  the sublevels $\gamma$ ($\gamma'$) in the Stark manifolds that correlate to a given $j$ ($j'$) state in the zero field limit.}
	\label{fig:LiFH_totalCS}
\end{figure}

Figure~\ref{fig:LiFH_totalCS} shows the total reaction cross section for the formation of HF($v'=0$) products   as a function of applied electric field. 
The LiFH PES features a significant activation barrier, which suppresses the tunneling of the heavy F atom, the main reaction mechanism   at low temperatures    \cite{Weck:05}, so the reaction cross section is small. An applied electric field causes  broad oscillations in the reaction cross section,  resulting from the field-induced coupling between the states of different total angular momenta \cite{Tscherbul:15}. These couplings open reaction pathways that are forbidden at zero field by the conservation of $J$.

As shown in Fig.~\ref{fig:LiFH_totalCS}, the inclusion of five TAM blocks in the present calculations ($J_\text{max}=4$) markedly improves the convergence of the reaction cross sections above 40 kV/cm. We found that in the $s$-wave limit, the number of TAM states required for convergence is approximately equal to the number of  rotational states of the reactant molecule coupled by the dc electric field. At fields below 40 kV/cm, the lowest pendular state of the LiF($v=1,{j}=0$)  reactant molecules is composed of 4 lowest bare rotational states of LiF, and hence adequate convergence is  achieved at $J_\text{max}=3$. At higher electric fields, the $j=4$ bare state of LiF becomes significantly admixed, necessitating the inclusion of the $J_\text{max}$ basis state for quantitatively accurate results.


Figure~\ref{fig:LiFH_totalCS}  also reveals narrow  field-induced reactive scattering resonances. These resonances are induced by couplings due to tunnelling through the reaction barrier and can therefore be tuned by an electric field that introduces a differential shift of the energies of the closed-channel quasi-bound states in the reactants arrangement. While a detailed analysis of these {\it field-induced reactive scattering resonances} is  beyond the scope of this work, we note that they  occur in the experimentally accessible range of electric fields (0 -- 200 kV/cm). 
It is important to emphasize, however, that both the positions and widths of the resonances are sensitive to the details of the PES \cite{Morita:19b}. Therefore, a theoretical analysis without input from experiments is likely insufficient to predict the positions of these resonances with quantitative accuracy, necessitating a close collaboration between theory and experiment.


Figure~\ref{fig:LiFH_totalCS} illustrates three important results. First, an external electric field can be used to tune scattering resonances that have a dramatic effect on the reactive scattering cross sections, even for reactions that occur by tunnelling of heavy atoms such as F. 
Second, an external electric field can have a significant effect on ultracold reactive scattering, even in the absence of scattering resonances, due to field-induced couplings between states of different total angular momenta. Third, it is evident that rigorous calculations of cross sections for reactions of polar molecules in electric fields require multiple (as many as four) total angular momentum states for convergence, even in the $s$-wave scattering regime. This indicates that high-order field-induced couplings have a significant effect on the reactive scattering cross sections.

\subsection{F~+~HD$(v=0,j=0)$ $\to$ HF~+~D, DF + H}

 
 The highly exothermic F~+~H$_2$ $\to$ HF~+~H chemical reaction and its isotopic variants has played a pivotal role in the development of modern theories of chemical reaction dynamics \cite{Althorpe:03,Clary:98,Skodje:00}.  
Recent experimental and theoretical studies have continued to explore this reaction through a combination of high-resolution crossed molecular beam experiments  and rigorous quantum dynamics calculations based on highly accurate {\it ab initio} PESs including non-adiabatic effects arising from the spin-orbit interaction in the open-shell F($^2$P) atom \cite{Alexander:00b,Chen:21}. Here, we  use the approach developed in Sec.~II to explore the effects of  external electric fields on the quantum dynamics of the F~+~HD $\to$ HD~+~D and DF~+~H reaction at low temperatures. We note that the F~+~H$_2$ reaction has already been studied experimentally at a temperature of  11~Kelvin \cite{Tizniti:14}. To maximize the possibility of observing interesting electric field effects, we choose to explore the collision energy of $E_C=1$~Kelvin, 
which corresponds to the lowest end of the collision energy range achievable in modern crossed-beam experiments with decelerated molecules \cite{Vogels:15}.

Our calculations on the F~+~HD reaction are based on the benchmark  Stark-Werner  (SW) PES \cite{Stark:96} provided as part of the standard ABC code \cite{Skouteris:00}. We performed extensive convergence tests  to get the state-to-state reaction probabilities converged to significantly better than 10\%.  The resulting basis contained all  FD  states  with rovibrational energies $E_{vj}\leq E_\text{max}=1.7$~eV,  $j\leq j_\text{max}= 29$, and $J\leq J_\text{max}=3$. We found that a large value of the rotational cutoff parameter $j_\text{max}$ is required to ensure convergence of the small state-resolved reaction probabilities into the final $v'=0$ manifold of DF and HF.  To this   end, we also used a fine hyperradial propagation grid  extending from 1.53~$a_0$ to 40~$a_0$ with 1,400 sectors.

We note that the  SW PES  is not the most accurate PES currently available for the F~+~HD chemical reaction. It is known to overestimate the F~+~H$_2$ reaction rate by a factor of $\simeq\,$3 at low collision energies \cite{Tizniti:14} and to offer only a partially adequate description of the transition state resonance in the HF product channel.
While more accurate F-HD PESs have been developed \cite{Sun:15} and shown to provide the reaction observables in closer agreement with experiment \cite{Sun:15,DeFazio:16}, the focus of this work is on determining the effects of external electric fields on reaction dynamics. Although our calculations can be performed on any PES, we use the benchmark SW PES here because it accurately describes the direct DF~+~H product channel (see Fig.~1 of Ref.~\citenum{DeFazio:16}), which is the most sensitive to the applied electric field.


\begin{figure}[t!]
	\centering
	\includegraphics[width=0.8\columnwidth, trim = 0 0 0 0]{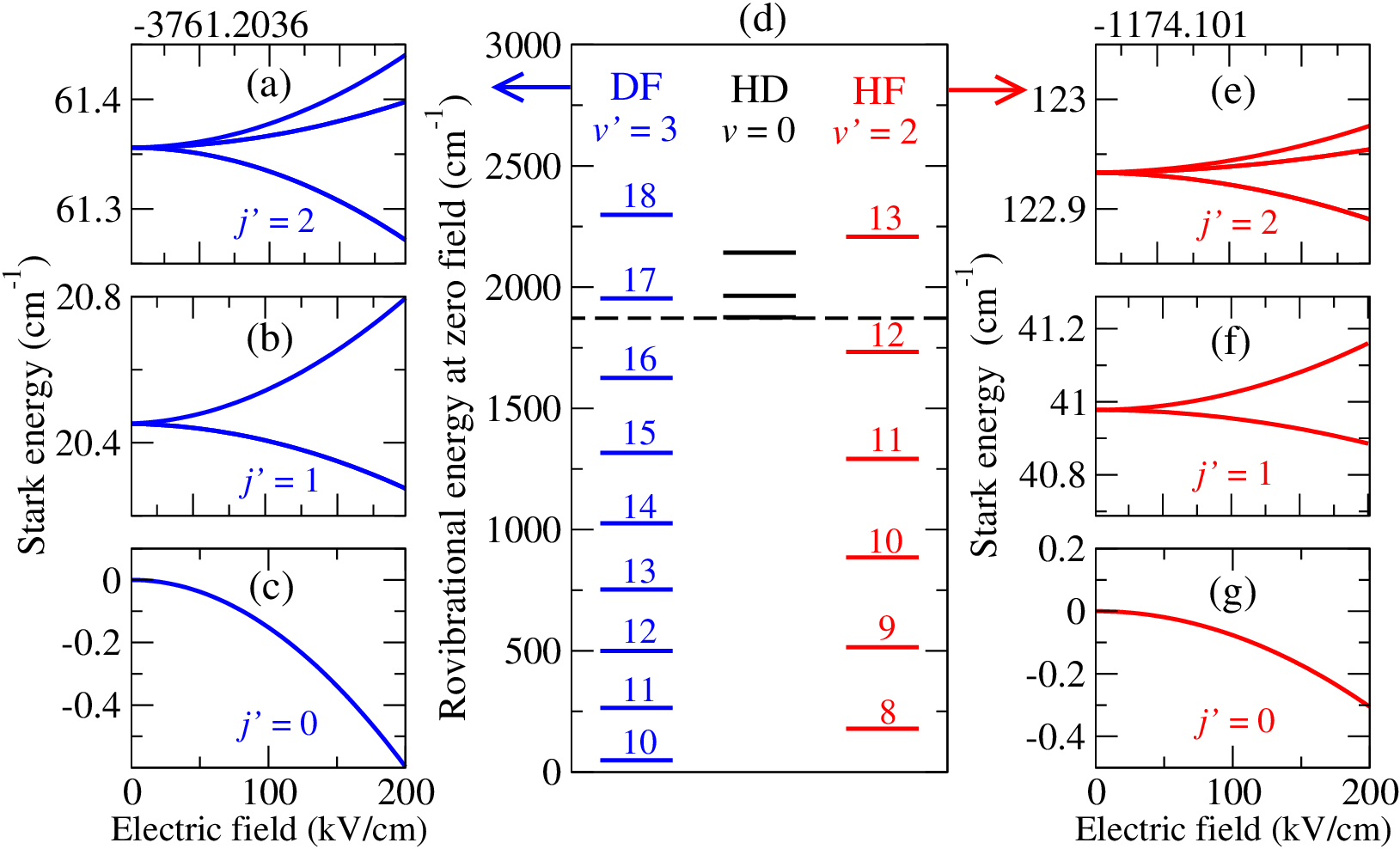}
	\caption{Energy level diagrams of the reactants and products of the chemical reaction F~+~HD$(v=1,j=0)$~ $\to$ HF~+~D, DF~+~H based on the Stark-Werner PES \cite{Stark:96}. }
	\label{fig:energeticsFHD}
\end{figure}

Figure \ref{fig:energeticsFHD} shows the  energy levels of the reactants and products for this reaction as a function of the electric field strength.  The interaction of the reactants with the electric field can be safely neglected due to the vanishinlgy small electric dipole moment of HD  ($5.85\times 10^{-4}$D) \cite{Trefler:68}.
The  HF and DF product molecules do have nonzero dipole moments (1.827 and 1.818 D, respectively), and hence experience substantial Stark shifts  in an electric field. The rotational constant of DF is approximately twice as small as that of HF, and hence, the Stark shift of DF($j=0$) is about a factor of two larger as shown in Fig.~\ref{fig:energeticsFHD}.  Because of the  large reaction exothermicity, a large number of rovibrational states of HF and DF   can be populated during the reaction.  Previous calculations have established that most of the populated states are rotationally excited, and hence have smaller Stark shifts than the lowest rotational states of HF and DF shown in Fig.~\ref{fig:energeticsFHD}. Indeed, the Stark shift of the $j$-th state scales to first order as $\simeq d^2/\Delta_{j,j-1}^2$, where $\Delta_{j,j-1}= 2 B_e j$ is the energy gap between the $j$-th and $(j-1)$-th states and $B_e$ is the rotational constant. Thus, the Stark shift  of the $j$-th rotational state decreases quadratically with $j$, and its magnitude for $j=10$ does not exceed 0.01 cm$^{-1}$ at $E=200$ kV/cm.

Based  on the energetic considerations,  we expect only a small effect of the electric field on the total  rate of the F~+~HF $\to$ DF~+~H, HF~+~D reaction at 1 K. Indeed,  our calculations  show that the electric field of 200 kV/cm  modifies the total reaction probabilities for either the HF or the DF product channels by $<$1~\%, which is consistent with zero, given the $<$5\%  convergence error of our calculations.

\begin{figure}[t!]
	\centering
	\includegraphics[width=0.5\columnwidth, trim = 0 0 0 0]{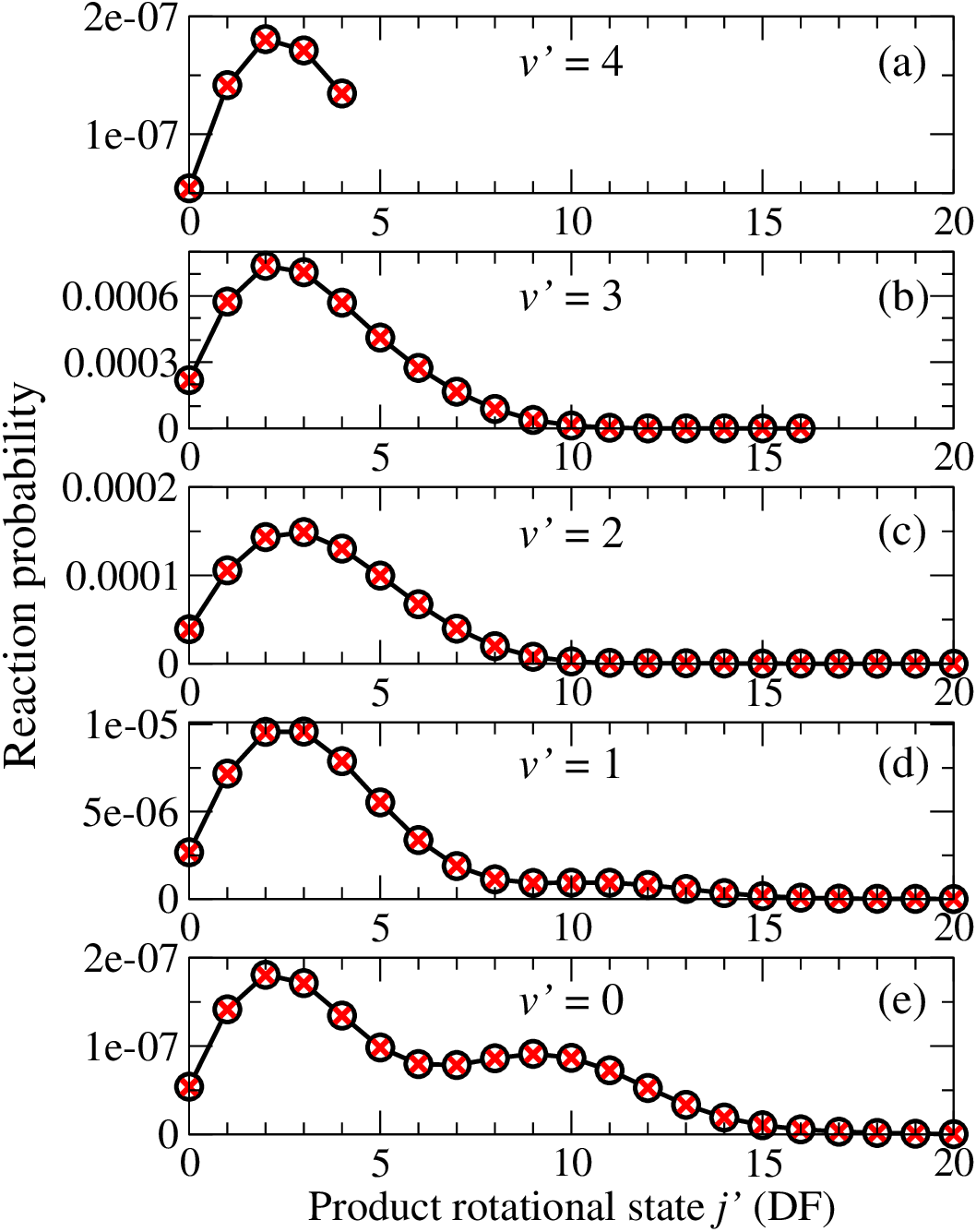}
	\caption{Rotational state distributions of DF$(v',j')$ products from the F~+~HD chemical reaction calculated at zero electric field (black circles) and at $E=200$~kV/cm (red crosses) for $v'=4$ (a), $v'=3$ (b), $v'=2$ (c), $v'=1$ (d), and $v'=0$ (e). The collision energy is 1~Kelvin.  }
	\label{fig:jdist_DF}
\end{figure}

\begin{figure}[t!]
	\centering
	\includegraphics[width=0.5\columnwidth, trim = 0 0 0 0]{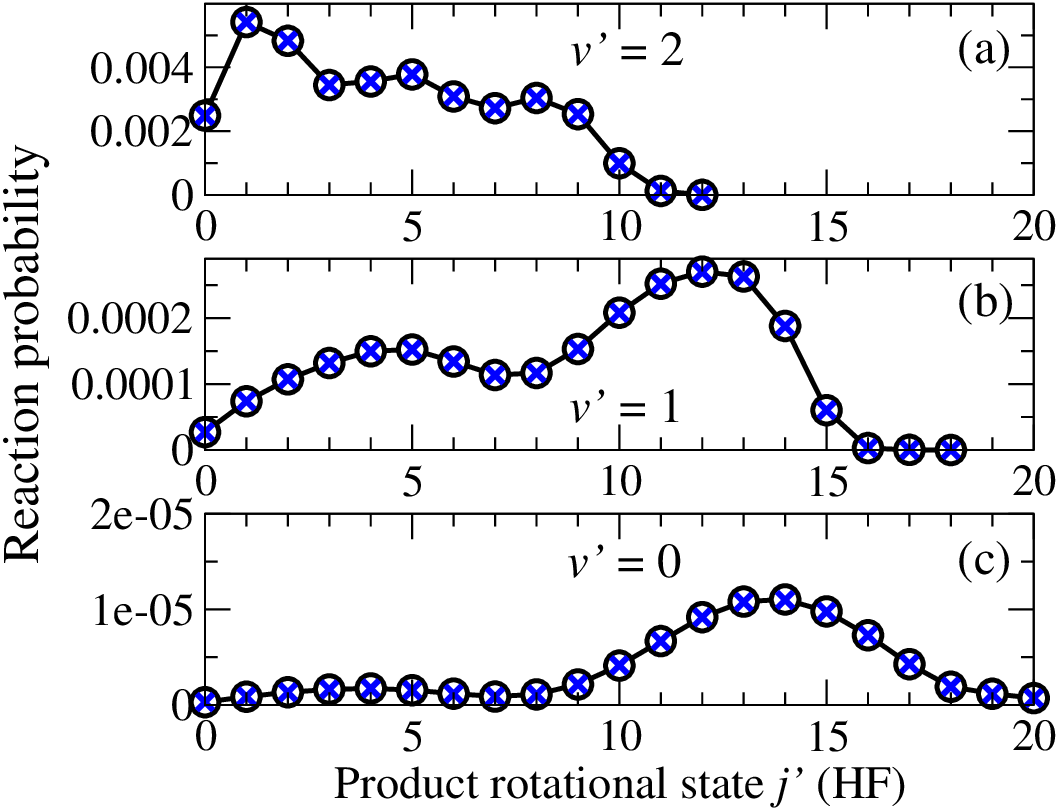}
	\caption{Rotational state distributions of HF$(v',j')$ products from the F~+~HD chemical reaction calculated at zero electric field (black circles) and at $E=200$~kV/cm (red crosses) for $v'=2$ (a), $v'=1$ (a), and $v'=0$. The collision energy is 1~Kelvin.  }
	\label{fig:jdist_HF}
\end{figure}

We next explore the effect of electric fields on {\it state-to-state}  reaction probabilities. In Fig.~\ref{fig:jdist_DF} we compare zero-field nascent rotational state distributions of the DF($v',j'$) products from the F~+~HD($v=0,j=0$) reaction with those at $E=200$ kV/cm.
  The distributions are nearly indistinguishable from each other, indicating that electric fields as high as 200 kV/cm have a negligible effect on the state-to-state F~+~HD $\to$ DF($v',j'$)~+~H reaction rates. The rotational state distributions of the HF($v',j'$) product channel shown in Fig.~\ref{fig:jdist_HF} are similarly insensitive to the electric field.
Interestingly, this remains true even for  the lowest rotational states  ($v', j'=0{-}2$) of DF and HF despite their appreciable Stark shifts compared to the collision energy (see Fig.~\ref{fig:energeticsFHD}). 

 This can be explained by noting that   the kinetic energy in the final  $j'=0$ product channels (hundreds of Kelvin) is much larger than the Stark shifts of these levels (less than 1~Kelvin). The  Stark effect therefore cannot alter the scattering wavefunction in the outgoing product channels of the strongly exothermic reaction.
Because the reactant channel is not affected by the field, the overall effect on the reaction rate is negligible. We expect this conclusion  to hold for all strongly exothermic chemical reactions with non-polar reactants.


\section{Conclusion}

We have presented a computationally efficient methodology for including the effects of external fields into rigorous quantum reactive scattering calculations. 
External fields break the spherical symmetry and increase the complexity of the numerical calculations by increasing the dimensionality of the computational Hilbert spaces. 
Our goal is to account for interactions of reactants and products with external fields at a minimal cost to the computation efficiency. To achieve this, 
the present formulation takes advantage of the total angular momentum (TAM) representation of molecular basis states.  
Although the total angular momentum is not conserved in the presence of an axially symmetric field, the Hamiltonian matrix is block-tridiagonal in the TAM representation. We have shown that this can be used to engineer an efficient basis set truncation that reduces the basis set size to a great extent.  We have also demonstrated an efficient implementation of the reactive scattering boundary conditions that significantly reduces the computation time of the reactive scattering calculations in the presence of fields. The present approach is a generalization of 
the formalism in Ref. \cite{Schatz:88,Pack:87}, based on the Fock-Delves hyperspherical coordinates. Our computation strategies, including the projections of the numerical solutions onto the field-dressed states of reactants and products, can be combined with other hyperspherical coordinates and thus applied to insertion as well as abstraction atom - molecule  reactions.

The speed-up afforded by the methodology demonstrated in this work enables rigorous computational studies of external field effects for a wide range of atom-diatom chemical reactions at low temperatures, including in the multiple partial wave scattering regime. This allowed us to extend our calculations of cross sections for two benchmark chemical reactions
to higher collision energies and higher angular momentum states than was previously feasible. The present calculations demonstrate that 
an external electric field can be used to modify reactive scattering cross sections both by tuning field-dependent scattering resonances, including resonances due to tunneling-driven interactions between the reactant and product chemical arrangements, and by coupling states of different total angular momenta, thus opening new reaction pathways.

\begin{figure}[t!]
	\centering
	\includegraphics[width=0.5\columnwidth, trim = 0 0 0 0]{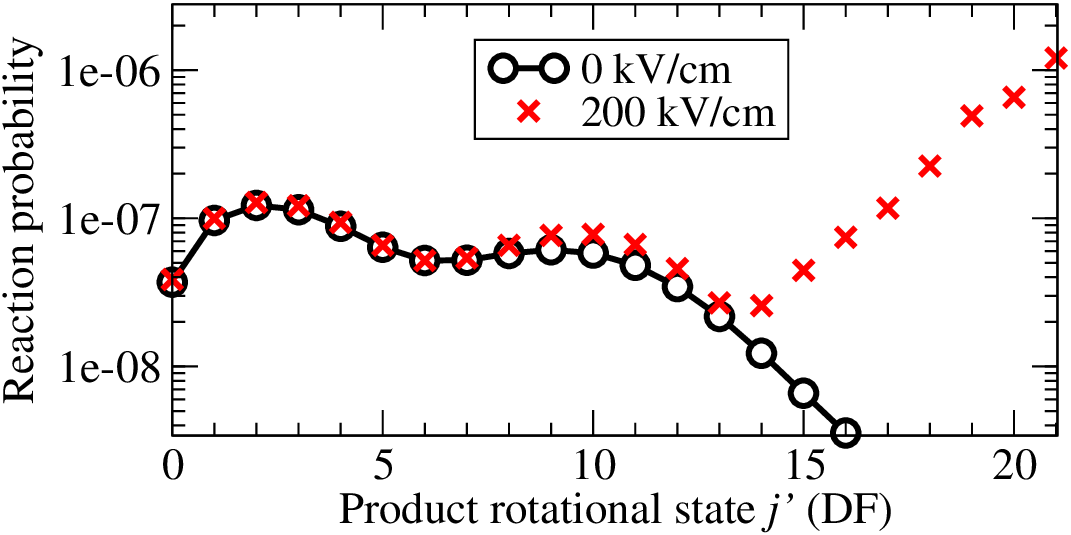}
	\caption{Reaction probabilities for the F~+~HD($v=0,j=0$) $\to$ DF$(v'=0,j')$~+~H plotted as a function of $j'$. Reaction probabilities at zero electric field are computed using the converged basis set (see text) whereas the  reaction probabilities at $E=200$ kV/cm are computed using a reduced basis set with $j_\text{max}=21$, $E_\text{max}=1.7$~eV, $N_\rho=1500$, and $\rho_\text{max}=50$~$a_0$. The probabilities are summed over $M=0$ and $M=\pm1$.
	 }
	\label{fig:unconverged}
\end{figure}

We have also shown that rigorous calculations of cross sections for reactions of polar molecules in electric fields require multiple total angular momentum states for convergence. It is particularly striking to observe that, even for reactions of molecules initially in the rotational state of zero angular momentum in the $s$-wave scattering regime, convergence requires as many as 5   
blocks of total angular momenta of the Hamiltonian matrix. This indicates that high-order field-induced couplings may have a significant effect on the reactive scattering cross sections.  
This also suggest that convergence of numerical calculations is especially important for interpreting reaction dynamics of molecules in the presence of external fields.
To illustrate this, we have repeated the calculations of Fig. \ref{fig:jdist_DF} (lower panel) with a reduced basis set. The results of these reduced-basis calculations for an electric field of 200 kV/cm, shown in Fig. \ref{fig:unconverged}, agree well with fully converged cross sections for field-free scattering producing molecules in low-energy rotational states, but
exhibit significant differences with field-free calculations for high-$j$ rotational states of the reaction products. These differences vanish as the basis set is increased to ensure convergence. The results of  Figs. \ref{fig:jdist_DF} and \ref{fig:unconverged} thus illustrate that spurious effects due to limited basis sets may appear as physical and that convergence of state-resolved cross sections, including for molecular states of high angular momentum, is essential for the proper interpretation of the effects of external fields on reaction dynamics.  
 This is in keeping with an earlier computational study of cold NH~+~NH collisions in a magnetic field, which found that both the elastic and inelastic cross sections (whose ratio is a crucial figure of merit for sympathetic cooling)  display a strong dependence on  rotational basis set size \cite{Suleimanov:16}.


Finally, we note that the present formalism can be applied to reactive scattering in the presence of either electric or magnetic fields. The present formulation can also be used to study reactions in the simultaneous presence of dc electric and dc magnetic fields, provided the field vectors are co-aligned. Our formalism can be readily extended to reactions in crossed electric and magnetic fields with a non-zero angle between the field vectors. In this case, the projection of the total angular momentum is not conserved \cite{Tscherbul:06,Abrahamsson:07}. However, the TAM representation and the methodology for the numerical application of the scattering boundary conditions can still be used as described in the present work. The numerical results of this work suggest that such calculations should be expected to require several $M$-states to be included simultaneously in the basis sets. This will increase the computation time to a great extent and it remains to be seen if rigorous calculations of cross sections for reactive scattering in crossed electric and magnetic fields are feasible with current computing hardware. 




\begin{acknowledgments}
This article is dedicated to the memory of our former advisor and colleague Dr. Alexei Buchachenko (1965--2023), 
 whose kindness, patience, and unwavering support throughout the  early stages of our careers will always be dearly remembered.
 Work at UNR was supported by the NSF CAREER program (grant No. PHY-2045681) and at UBC by NSERC of Canada. 
\end{acknowledgments}



\newpage
\appendix
\section{Boundary conditions in hyperspherical coordinates in the presence of an external electric field}

Here, we derive the expressions for the hyperradial solution matrix $\bm{\Gamma} (\rho)$  and its hyperradial derivative  $\frac{\partial\bm{\Gamma}(\rho)}{\partial\rho}$  given by Eqs.~\eqref{Gamma_ni} and \eqref{matrix_dGamma} of Sec. \ref{sec:BC}.

\subsection{Derivation of Eq.~\eqref{Gamma_ni}  of the main text}

Our starting point is Eq.~\eqref{starting_point}
 \begin{equation}\label{starting_point_app}
 \Gamma_{ni}(\rho)  = \langle n | \rho^{5/2} \Psi^{i}  \rangle_\text{FD},
 \end{equation}
where the subscripts ``FD'' and ``Jac'' are used to emphasize the distinction between the basis functions in the FD and Jacobi coordinates, which is already  encoded in state indices (see Table I).

Substituting the asymptotic expression for  $\Psi^i$ in Jacobi coordinates \eqref{Jac_expansion} into Eq.~\eqref{starting_point_app}, expanding the field-dressed Jacobi and FD basis states  using Eqs.~\eqref{m_Jac} and \eqref{n_FD}, and noting that $F^{i}_{m}(R_{\alpha_m})= F_{mi}(R_{\alpha_m})$, we find 
 \begin{align}\notag
 \Gamma_{ni}(\rho)  &= {}_\text{FD} \langle n |  \sum_{m} \left(\frac{\rho^{5/2} }{R_{\alpha_m} r_{\alpha_m}} \right) F_{mi} (R_{\alpha_m})   \sum_{m'} C^\text{Jac}_{m' m}(E) |m'\rangle_\text{Jac} \\ \label{interm_eq1_app}
  &= \sum_{n',m'} C_{n'n}^\text{FD}(E) C^\text{Jac}_{m'm}(E) \sum_m   {}_\text{FD} \langle n' |  \Bigl{(} \frac{\rho^{5/2} }{R_{\alpha_m} r_{\alpha_m}}  \Bigr{)} F_{mi} (R_{\alpha_m})   |m'\rangle_\text{Jac}.
 \end{align}
 These expressions make explicit the field dependence of the matrix elements, which enters through the  Stark mixing coefficients  $C_{n'n}^\text{FD}(E)$ and $C^\text{Jac}_{m'm}(E)$. The matrix elements ${}_\text{FD} \langle n' | \left( \frac{\rho^{5/2} }{R_{\alpha_m} r_{\alpha_m}}  \right)  F_{mi} (R_{\alpha_m})   |m'\rangle_\text{Jac}$ are expressed in the field-free basis, and can be dealt with using the approach of Pack and Parker \cite{Pack:87}.
 
Specifically, on plugging in the expressions for  $ |m'\rangle_\text{Jac}$ and ${}_\text{FD} \langle n'|$ and noting that ${\rho^{5/2}}/(R_{\alpha_m} r_{\alpha_m}) = {2\rho^{1/2} }/{\sin{2\theta_{\alpha_m}}}$ (which follows from the definition of mass-scaled Jacobi cordinates \cite{Pack:87}), 
  the field-free matrix element  in Eq.~\eqref{interm_eq1_app} becomes  \cite{Pack:87}
  \begin{multline}\label{mxel_field_free_app}
  \langle n' |  \Bigl{(} \frac{\rho^{5/2} }{R_{\alpha_m} r_{\alpha_m}}  \Bigr{)}  F_{mi} (R_{\alpha_m})   |m'\rangle_\text{Jac} = \frac{1}{4} \int d\theta_{\alpha_{n'}}  \sin^2\theta_{\alpha_{n'}} 
  \left[ \frac{2\chi_{\alpha_{n'} v_{n'}j_{n'}}(\theta_{\alpha_{n'}};\rho)}{\sin 2\theta_{\alpha_{n'}}}  \right]
\left[ \frac{2\rho^{1/2} }{\sin{2\theta_{\alpha_m}}}\right]  
\\ \times 
F_{mi} (R_{\alpha_m})  \xi_{\alpha_{m'} v_{m'} j_{m'}} (r_{\alpha_{m'}})
  \iint d\hat{R}_{\alpha_{n'}} d\hat{r}_{\alpha_{n'}} \mathcal{J}^{J_{n'}M}_{j_{n'} l_{n'}} (\hat{R}_{\alpha_{n'}},\hat{r}_{\alpha_{n'}})
   \mathcal{J}^{J_{m'}M}_{j_{m'} l_{m'}} (\hat{R}_{\alpha_{m'}},\hat{r}_{\alpha_{m'}})
\end{multline}
Using the decoupling between the different arrangements in the limit of large $\rho$ (which gives the factor $\delta_{\alpha_{n'}\alpha_{m'}}$)
and the orthonormality of bipolar spherical harmonics within a single arrangement
  \begin{equation}\label{sph_harmonics_orthog_app}
   \iint d\hat{R}_{\alpha_{n'}} d\hat{r}_{\alpha_{n'}} \mathcal{J}^{J_{n'}M}_{j_{n'} l_{n'}} (\hat{R}_{\alpha_{n'}},\hat{r}_{\alpha_{n'}})
   \mathcal{J}^{J_{m'}M}_{j_{m'} l_{m'}} (\hat{R}_{\alpha_{m'}},\hat{r}_{\alpha_{m'}})
= \delta_{J_{n'}J_{m'}}\delta_{j_{n'}j_{m'}}\delta_{l_{n'}l_{m'}},
  \end{equation}
 we obtain the field-free matrix element from Eq.~\eqref{mxel_field_free_app} as
  \begin{equation}\label{mxel_field_free2_app}
    \langle n' |  \Bigl{(} \frac{\rho^{5/2} }{R_{\alpha_m} r_{\alpha_m}}  \Bigr{)}  F_{mi} (R_{\alpha_m})   |m'\rangle_\text{Jac}
     =  \int d\theta_{\alpha_{m'}} X_{n'm'} (\theta_{\alpha_{m'}},r_{\alpha_{m'}};\rho) F_{mi}(R_{\alpha_{m'}}),
     \end{equation}
 where
    \begin{equation}\label{Xprimed_app}
  X^{(0)}_{n'm'} (\theta_{\alpha_{m'}},r_{\alpha_{m'}};\rho) = \rho^{1/2}  \delta_{\alpha_{n'}\alpha_{m'}} \delta_{J_{n'}J_{m'}} \delta_{j_{n'}j_{m'}}\delta_{l_{n'}l_{m'}} 
  \chi_{\alpha_{n'} v_{n'}j_{n'}}(\theta_{\alpha_{n'}};\rho)  \xi_{\alpha_{m'} v_{m'} j_{m'}} (r_{\alpha_{m'}}).
     \end{equation}
  Combining these results with Eq.~\eqref{interm_eq1_app}, we obtain
  \begin{equation}\label{Gamma_ni_app}
\Gamma_{ni} (\rho) =  \sum_m \int  X_{nm}(\theta_{\alpha_{m}},r_{\alpha_{m}};\rho) F_{mi}(R_{\alpha_{m}})d\theta_{\alpha_{m}},
\end{equation}
where the  hyperangle-dependent overlap matrix is obtained by transforming the field-free overlap matrix \eqref{Xprimed_app} to the field-dressed basis 
  \begin{equation}\label{X_nm_app}
X_{nm}(\theta_{\alpha_{m}},r_{\alpha_{m}};\rho)  = \sum_{n',m'}C_{n'n}^\text{FD}(E)     X^{(0)}_{n'm'}(\theta_{\alpha_{m'}},r_{\alpha_{m'}};\rho) C_{m'm}^\text{Jac}(E).
\end{equation}
    This completes the derivation of Eq.~\eqref{Gamma_ni}  of the main text.

\subsection{Derivation of Eq.~\eqref{matrix_dGamma}  of the main text}
 
The derivative of the hyperradial solution matrix  $\bm{\Gamma}(\rho)$, which is required to impose the boundary conditions on the log-derivative matrix, is readily obtained by differentiating Eq.~\eqref{Gamma_ni_K} of the main text with respect to $\rho$
\begin{equation}\label{Gamma_ni_K_app}
\frac{\partial\Gamma_{ni} (\rho)}{\partial\rho} = 
\int d\theta_{\alpha_{i}}  \frac{\partial }{\partial\rho} \bigl{\{} X_{ni}(\theta_{\alpha_{i}},r_{\alpha_{i}};\rho) a_{ii}(R_{\alpha_{i}})\bigr{\}}
- \sum_m K_{mi}  \int d\theta_{\alpha_{i}} 
 \frac{\partial }{\partial\rho} \bigl{\{}  X_{nm}(\theta_{\alpha_{m}},r_{\alpha_{m}};\rho) b_{mm}(R_{\alpha_{m}})\bigr{\}}.
\end{equation}
The integrand in the second  term on the right-hand side may be written as  
\begin{equation}\label{integrand_diff_app}
 \frac{\partial }{\partial\rho} \bigl{\{}  \rho^{1/2}\tilde{X}_{nm}(\theta_{\alpha_{m}},r_{\alpha_{m}};\rho) b_{mm}(R_{\alpha_{m}})\bigr{\}},
\end{equation}
where the scaled matrix elements [cf. Eqs.~\eqref{X_nm}]
\begin{equation}\label{X_tilde_app}
\tilde{X}_{nm}(\theta_{\alpha_{m}},r_{\alpha_{m}};\rho) =  \rho^{-1/2} {X}_{nm}(\theta_{\alpha_{m}},r_{\alpha_{m}};\rho)  =  \sum_{n',m'}C_{n'n}^\text{FD}(E)     \tilde{X}^{(0)}_{n'm'}(\theta_{\alpha_{m'}},r_{\alpha_{m'}};\rho) C_{m'm}^\text{Jac}(E)
\end{equation}
with 
  \begin{equation}\label{X_npmp_app}
   \tilde{X}^{(0)}_{n'm'}(\theta_{\alpha_{m'}},r_{\alpha_{m'}};\rho) =
   \delta_{\alpha_{n'}\alpha_{m'}} \delta_{J_{n'} J_{m'}}\ \delta_{j_{n'} j_{m'}}\delta_{l_{n'} l_{m'}}
 \chi_{\alpha_{n'}v_{n'}j_{n'}}(\theta_{\alpha_{n'}};\rho)   \xi_{\alpha_{m'} v_{m'} j_{m'}} (r_{\alpha_{m'}}) 
  \end{equation}
The only difference between these scaled matrix elements  and those defined in Eqs.~\eqref{X_nm} and \eqref{X_npmp} is the absence of  the prefactor $\rho^{1/2}$. 
The advantage of Eqs.~\eqref{X_tilde_app} and \eqref{X_npmp_app} is that they facilitate the evaluation of hyperradial derivatives of matrix elements. Indeed,  Eq.~\eqref{integrand_diff_app} may be written as
\begin{equation}\label{integrand_diff_app2}
\frac{1}{2}\rho^{-1/2} \tilde{X}_{nm}(\theta_{\alpha_{m}},r_{\alpha_{m}};\rho) b_{mm}(R_{\alpha_m})
+ \rho^{1/2} \frac{\partial }{\partial\rho} \bigl{\{}  \tilde{X}_{nm}(\theta_{\alpha_{m}},r_{\alpha_{m}};\rho) b_{mm}(R_{\alpha_{m}})\bigr{\}}.
\end{equation}
The product matrix elements $\tilde{X}_{nm}(\theta_{\alpha_{m}},r_{\alpha_{m}};\rho)$ bear explicit dependence on $\rho$ only via the arguments of the FD basis functions. This dependence can be neglected within a given propagation sector (here, we assume that  $\rho=\rho_a$ lies in the last sector).  By contrast, the ${X}_{nm}(\theta_{\alpha_{m}},r_{\alpha_{m}};\rho)$] do explicitly depend on $\rho$ via the prefactor $\rho^{1/2}$.

 Using the chain rule, the hyperradial derivative term in Eq.~\eqref{integrand_diff_app2} evaluates to
\begin{align}\label{integrand_diff_app3}\notag
\frac{\partial }{\partial\rho} \bigl{\{}  \tilde{X}_{nm}(\theta_{\alpha_{m}},r_{\alpha_{m}};\rho) b_{mm}(R_{\alpha_{m}})\bigr{\}} &=
\cos \theta_{\alpha_n}   \frac{\partial}{\partial R_{\alpha_n}}  \bigl{\{}  \tilde{X}_{nm}(\theta_{\alpha_{m}},r_{\alpha_{m}};\rho) b_{mm}(R_{\alpha_{m}})\bigr{\}} 
\\
  &+     \sin\theta_{\alpha_n}\frac{\partial}{\partial r_{\alpha_n}}  \bigl{\{}  \tilde{X}_{nm}(\theta_{\alpha_{m}},r_{\alpha_{m}};\rho) b_{mm}(R_{\alpha_{m}})\bigr{\}},
\end{align}
where we have used the relations $\partial R_{\alpha_n}/\partial \rho=\cos \theta_{\alpha_n}$ and $\partial r_{\alpha_n}/\partial \rho=\sin \theta_{\alpha_n}$ which follow immediately  from the definition of the FD coordinates  \cite{Pack:87}, $R_{\alpha_n}=\rho\cos \theta_{\alpha_n} $ and $r_{\alpha_n}=\rho\sin \theta_{\alpha_n}$.

Using $X_{nm}(\theta_{\alpha_{m}},r_{\alpha_{m}};\rho) = \rho^{1/2}\tilde{X}_{nm}(\theta_{\alpha_{m}},r_{\alpha_{m}};\rho) $ [see Eq.~\eqref{integrand_diff_app}] and taking the radial derivatives in Eq.~\eqref{integrand_diff_app3} we can rewrite Eq.~\eqref{integrand_diff_app2} as
 \begin{multline}\label{integrand_diff_app4}
\frac{1}{2\rho} {X}_{nm}(\theta_{\alpha_{m}},r_{\alpha_{m}};\rho) b_{mm}(R_{\alpha_m}) \\
+ \rho^{1/2} \biggl{[}
\cos \theta_{\alpha_n}   \tilde{X}_{nm}(\theta_{\alpha_{m}},r_{\alpha_{m}};\rho)  \frac{\partial b_{mm}(R_{\alpha_{m}})}{\partial R_{\alpha_n}} 
  +     \sin\theta_{\alpha_n}  b_{mm}(R_{\alpha_{m}}) \frac{\partial \tilde{X}_{nm}(\theta_{\alpha_{m}},r_{\alpha_{m}};\rho)}{\partial r_{\alpha_n}} \biggr{]}.
\end{multline}
 From this, we obtain  the second term on the right-hand side of Eq.~\eqref{Gamma_ni_K_app}  as
\begin{multline}\label{Gamma_ni_K_app_2nd_term}
 -\sum_m  K_{mi} \biggl{\{} \frac{1}{2\rho} \int d\theta_{\alpha_{m}} X_{nm}(\theta_{\alpha_{m}},r_{\alpha_{m}};\rho) b_{mm}(R_{\alpha_m})  
\\
 - \int d\theta_{\alpha_{n}}
 \cos \theta_{\alpha_n} \rho^{1/2} \tilde{X}_{nm}(\theta_{\alpha_{m}},r_{\alpha_{m}};\rho)   \frac{\partial b_{mm}(R_{\alpha_{m}})}{\partial R_{\alpha_n}}  
+     \sin\theta_{\alpha_n} b_{mm}(R_{\alpha_{m}}) \rho^{1/2} \frac{\partial \tilde{X}_{nm}(\theta_{\alpha_{m}},r_{\alpha_{m}};\rho)}{\partial r_{\alpha_n}}
  \biggr{\}}
\end{multline}
or, in matrix form, as the $(n,i)$-th element of the matrix product $-\left(\frac{1}{2\rho}  \mathbf{B} + \mathbf{H} \right) \mathbf{K}$
\begin{equation}\label{Gamma_ni_K_app_2nd_term_mx}
  -\sum_m \left( \frac{1}{2\rho}  B_{nm} + H_{nm}\right) {K}_{mi},
\end{equation}
where the matrix elements $B_{nm}$ are given by Eq.~\eqref{A_and_B} and
\begin{equation}\label{G_app}
{H}_{nm} = \rho^{1/2}    \int d\theta_{\alpha_n}
  \biggl{[}  \cos\theta_{\alpha_n}  \frac{\partial b_{mm}(R_{\alpha_n})}{\partial R_{\alpha_n}} \tilde{X}_{nm}(\theta_{\alpha_{n}},r_{\alpha_{i}};\rho) 
  +
 \sin\theta_{\alpha_n} b_{mm}(R_{\alpha_n})  \frac{\partial \tilde{X}_{ni}(\theta_{\alpha_{i}},r_{\alpha_{n}};\rho) }{\partial r_{\alpha_n}}   
    \biggr{]}.
\end{equation}
The derivative with respect to $r_{\alpha_n}$ can be readily evaluated from Eq.~\eqref{X_tilde_app} since the electric field-mixing amplitudes  $C_{n'n}^\text{FD}(E)$ and $C_{m'm}^\text{Jac}(E)$ are independent of $\rho$ within a given sector
\begin{equation}\label{X_tilde_deriv_app}
\frac{\partial \tilde{X}_{nm}(\theta_{\alpha_{m}},r_{\alpha_{m}};\rho)}{\partial r_{\alpha_n}} 
=\sum_{n',m'}C_{n'n}^\text{FD}(E)   
\frac{\partial  \tilde{X}^{(0)}_{n'm'}(\theta_{\alpha_{m'}},r_{\alpha_{m'}};\rho)  }{\partial r_{\alpha_n}}
  C_{m'm}^\text{Jac}(E).
\end{equation}
The derivative on the right-hand side is obtained by differentiating Eq.~\eqref{X_npmp_app} with respect to $r_{\alpha_n}$ (note that $\alpha_m=\alpha_n=\alpha_{n'}=\alpha_{m'}$ because reactive scattering boundary conditions are applied in the asymptotic region $\rho\to \infty$, where the different reaction arrangements are completely decoupled, even in the presence of electric fields)
\begin{equation}\label{X_tilde_deriv2_app}
\frac{\partial  \tilde{X}^{(0)}_{n'm'}(\theta_{\alpha_{m'}},r_{\alpha_{m'}};\rho)  }{\partial r_{\alpha_{m'}}} 
    =
   \delta_{\alpha_{n'}\alpha_{m'}} \delta_{J_{n'} J_{m'}}\ \delta_{j_{n'} j_{m'}}\delta_{l_{n'} l_{m'}}
 \chi_{\alpha_{n'}v_{n'}j_{n'}}(\theta_{\alpha_{n'}};\rho)  \frac{\partial{\xi_{\alpha_{m'} v_{m'} j_{m'}}}(r_{\alpha_{m'}})}{\partial r_{\alpha_{m'}}}.
  \end{equation}

To complete the derivation, we finally consider the first term in Eq.~\eqref{Gamma_ni_K_app}, which may be written as
\begin{multline}\label{Gamma_ni_K_app2}
\int d\theta_{\alpha_{n}}  \frac{\partial }{\partial\rho} \bigl{\{} \rho^{1/2} \tilde{X}_{ni}(\theta_{\alpha_{n}},r_{\alpha_{n}};\rho) a_{ii}(R_{\alpha_{n}})\bigr{\}} \\ = 
\int d\theta_{\alpha_{n}}  \frac{1}{2\rho^{1/2}}  \tilde{X}_{ni}(\theta_{\alpha_{n}},r_{\alpha_{n}};\rho) a_{ii}(R_{\alpha_{n}}) 
+\int d\theta_{\alpha_{n}} \rho^{1/2} \frac{\partial}{\partial \rho} \bigl{\{}  \tilde{X}_{ni}(\theta_{\alpha_{n}},r_{\alpha_{n}};\rho) a_{ii}(R_{\alpha_{n}})\bigr{\}}
\end{multline}

Using the chain rule to evaluate the hyperradial derivative in the second term on the right-hand side (see above), we find
\begin{multline}\label{Gamma_ni_K_app3}
 \frac{1}{2\rho}  \int d\theta_{\alpha_{n}}  {X}_{ni}(\theta_{\alpha_{n}},r_{\alpha_{n}};\rho) a_{ii}(R_{\alpha_{n}}) 
\\ +  \rho^{1/2} \int d\theta_{\alpha_{n}} 
\biggl{[} \cos \theta_{\alpha_n}   \tilde{X}_{ni}(\theta_{\alpha_{n}},r_{\alpha_{n}};\rho)  \frac{\partial a_{ii}(R_{\alpha_{n}})}{\partial R_{\alpha_n}} 
  +     \sin\theta_{\alpha_n}  a_{ii}(R_{\alpha_{n}}) \frac{\partial \tilde{X}_{ni}(\theta_{\alpha_{n}},r_{\alpha_{n}};\rho)}{\partial r_{\alpha_n}} \biggr{]}.
\end{multline}

The first term in Eq.~\eqref{Gamma_ni_K_app} can thus be recast as $  \frac{1}{2\rho}  {A}_{ni} + G_{ni}$ [cf. Eq.~\eqref{Gamma_ni_K_app_2nd_term_mx}] or,  in matrix form,
\begin{equation}\label{Gamma_ni_K_app_2nd_term_mx2}
   \frac{1}{2\rho}  \mathbf{A} + \mathbf{G},
\end{equation}
where the matrix elements $A_{ni}$ are given by Eq.~\eqref{A_and_B} and 
\begin{equation}\label{G_app}
{G}_{ni} = \rho^{1/2}    \int d\theta_{\alpha_n}
  \biggl{[}  \cos\theta_{\alpha_n}  \frac{\partial a_{ii}(R_{\alpha_n})}{\partial R_{\alpha_n}} \tilde{X}_{ni}(\theta_{\alpha_{n}},r_{\alpha_{i}};\rho) 
  +
 \sin\theta_{\alpha_n} a_{ii}(R_{\alpha_n})  \frac{\partial \tilde{X}_{ni}(\theta_{\alpha_{i}},r_{\alpha_{n}};\rho) }{\partial r_{\alpha_n}}   
    \biggr{]}.
\end{equation}

Putting together the first and the second terms in Eq.~\eqref{Gamma_ni_K_app} and using $\bm{\Gamma}(\rho)= \mathbf{A} - \mathbf{B} \mathbf{K}  $, we find
\begin{equation}\label{Gamma_ni_K_app_2nd_term_mx2}
\frac{\partial\bm{\Gamma}(\rho)}{\partial\rho} =  \frac{1}{2\rho}  \mathbf{A} + \mathbf{G} - \left(  \frac{1}{2\rho}  \mathbf{B} + \mathbf{H} \right)\mathbf{K}
= \frac{1}{2\rho} \bm{\Gamma}(\rho) +  ( \mathbf{G}  -  \mathbf{H} \mathbf{K} ),
\end{equation}
completing the derivation of Eq.~\eqref{matrix_dGamma}  of the main text.



%
%

%


\bibliography{references.bib}

\end{document}